\documentclass[twocolumn]{aastex631}

\usepackage[T1]{fontenc}
\usepackage[utf8]{inputenc}
\usepackage{graphicx}
\usepackage{amssymb}
\usepackage{amsmath}
\usepackage{natbib}
\usepackage[capitalise]{cleveref}
\usepackage{appendix}
\usepackage{booktabs}
\usepackage{multirow}
\usepackage{array}
\newcolumntype{P}[1]{>{\centering\arraybackslash}p{#1}}
\usepackage{color}

\def\red#1{\textcolor{red}{\textbf{}}}
\def\blue#1{#1}

\begin{document}

\title{Late-Time Evolution of Magnetized Disks in Tidal Disruption Events}

\author{Yael Alush}
\affiliation{Racah Institute of Physics, The Hebrew University, 91904, Jerusalem, Israel}
\email{yael.alush@mail.huji.ac.il}

\author{Nicholas C. Stone}
\affiliation{Racah Institute of Physics, The Hebrew University, 91904, Jerusalem, Israel}
\affiliation{Department of Astronomy, University of Wisconsin, Madison, WI 53706, USA}

\begin{abstract}
In classic time-dependent 1D accretion disk models, the inner radiation pressure dominated regime is viscously unstable. However, late-time observations of accretion disks formed in tidal disruption events (TDEs) do not exhibit evidence of such instabilities. The common theoretical response is to modify the viscosity parametrization, but typically used viscosity parametrization are generally {\it ad hoc}. In this study, we take a different approach, and investigate a time-dependent 1D $\alpha$-disk model in which the pressure is dominated by magnetic fields rather than photons. We compare the time evolution of thermally stable, strongly magnetized TDE disks to the simpler linear viscosity model. We find that the light curves of magnetized disks evolve as $L_{\rm UV}\propto t^{-5/6}$ for decades to centuries, and that this same evolution can be reproduced by the linear viscosity model for specific parameter choices. Additionally, we show that TDEs remain UV-bright for many years, suggesting we could possibly find fossil TDEs decades after their bursts. We estimate that \textit{ULTRASAT} could detect hundreds of such events, providing an opportunity to study late-stage TDE physics and supermassive black hole (SMBH) properties. Finally, we explore the connection between TDE disks and quasi-periodic eruptions (QPEs) suggested by recent observations. One theoretical explanation involves TDE disks expanding to interact with extreme mass ratio inspirals (EMRIs), which produce X-ray flares as the EMRI passes through the disk. Our estimates indicate that magnetized TDE disks should exhibit QPEs earlier than those observed in AT2019qiz, suggesting that the QPEs may have begun before their first detection.
\end{abstract}

\section{Introduction}
\label{sec: intro}
Thermal and viscous instabilities are common features in the inner regions of simple accretion disk models around black holes (BHs) \citep{Pringle+1973,Shakura&Sunyaev1976,Lightman1974,Piran1978}. The inner zones of thin $\alpha$-disk models \citep{Shakura&Sunyaev1973} can be describe by three states. Radiation pressure often dominates over the gas pressure in the inner regions, but when the mass accretion rate is very low, radiation pressure becomes inefficient and the disk is stable \citep{Shakura&Sunyaev1976,Shen&Matzner2014}. Alternatively, the disk is also stable at very high accretion rates when advection is the dominant cooling mechanism \citep{Abramowicz1988+,Narayan&Yi1995}. However, at moderate accretion rates, where radiation pressure and radiative cooling dominate, the disk becomes thermally and viscously unstable. These instabilities are thought to produce global limit-cycle behavior, where the mass accretion rate alternates between the high-accretion-rate, advection-dominated branch and the low-accretion-rate, gas-pressure-dominated state \citep{Honma+1991,Szuszkiewicz&Miller1998,Janiuk+2002,Ohsuga2006,Shen&Matzner2014,Piro&Mockler2024}. 

Despite the theoretical prediction of instabilities in radiatively dominated regions, most observed X-ray binaries (XRBs) and active galactic nuclei (AGNs) appear stable. Evidence of instability has been observed in some cases, such as the `heartbeat' variability in XRBs, a series of bursts with recurrence times ranging from seconds to minutes \citep{Belloni+1997,Taam+1997,Massaro+2010,Altamirano+2011,Bagnoli+2015,Maselli+2018}, and the long-term variability of changing-look AGNs on timescales of weeks to months \citep{Matt+2013,Bianchi+2005,LaMassa2015,Ruan+2016,Lawrence2018}. While these variabilities may be linked to radiation pressure instabilities, most observed XRBs and AGNs remain stable, which challenges the standard $\alpha$-disk predictions \citep{Done+2007}.
 
Another accretion disk phenomenon that exhibits surprising stability is tidal disruption events (TDEs). A TDE occurs when a star approaches sufficiently close to a supermassive black hole (SMBH) and is torn apart by the SMBH's strong tidal forces \citep{Hills1975,Rees1988}. A portion of the material from the disrupted star is bound to the SMBH, while the other portion is unbound and escapes from the SMBH. Although the early stages of a TDE are complex \citep{Hayasaki+2013,Shiokawa+2015,Hayasaki+2016,Bonnerot+2020,Bonnerot&Stone2021}, in the later stages, the gas dissipates its excess energy and circularizes to form an accretion disk \citep{vanVelzen+2019,Steinberg&Stone2024}.

TDEs have high enough mass accretion rates such that radiation pressure would dominate gas pressure \citep{Ulmer1999,Wen+2020}. Once the accretion rate falls below the stable advection-dominated state, a 1D $\alpha$-model would predict that the disk should experience global limit-cycle behavior \citep{Shen&Matzner2014, Piro&Mockler2024}. However, observations of late-time TDEs show that the ultraviolet (UV) light curves flatten \citep{Gezari+2015, vanVelzen+2019} and generally remain quite constant \citep{Mummery+2024}. The observed UV luminosities are too bright to be consistent with the lower, gas-pressure-dominated branch of solutions \citep{vanVelzen+2019}, which can be orders of magnitude dimmer \citep{Shen&Matzner2014, Piro&Mockler2024}. Therefore, the standard $\alpha$-disk model alone does not explain the late-time TDE observations, indicating that something\footnote{While TDE disks are expected to be geometrically complex at early times, with nontrivial tilts \citep{Stone&Loeb2012, Franchini+2016}, wind losses \citep{Miller2015}, and global eccentricities \citep{Shiokawa+2015, Zanazzi+20}, we are concerned in this paper with very late time observations years post-disruption, at which point the simpler axisymmetric geometry of Shakura-Sunyaev type models should be applicable.} prevents thermal and viscous instabilities from developing in these disks.

A common theoretical solution to stabilize the disk involves changing the viscosity parametrization \citep{Sakimoto&Coroniti1981,Taam&Lin1984,Mummery2019}. However, commonly used viscosity parametrizations are ad hoc and lack physical motivation. It is likely that a more fundamental piece of physics is missing.

Several mechanisms have been proposed to explain the observed stability of most accretion disks, including magnetic fields \citep{Begelman&Pringle2007,Oda+2009}, viscous fluctuations in a turbulent flow\footnote{This mechanism appeals to properties of 3D turbulence.  Viscous instabilities have not been extensively studied in 3D MHD simulations because they operate on too long a timescale to easily simulate. However, 1D calculations suggest that thermal and viscous instabilities occur under the same conditions \citep{Piran1978}, and thermal instabilities have been successfully simulated in 3D MHD \citep{Jiang+2016, Mishra+2022}, confirming that they are not merely artifacts of 1D or $\alpha$-viscosity models.} \citep{Janiuk&Misra2012}, delayed pressure responses to stress variations in magnetized disks \citep{Hirose+2009}, the ``iron opacity bump'' \citep{Jiang+2016, Grzedzielski+2017}, and strong star-disk collisions  \citep{Linial&Metzger2024}. Among these, stabilization by magnetic fields seems to be the most generic.

Magnetohydrodynamic (MHD) simulations of tidally disrupted stars indicate that the resulting disk's magnetic field configuration is predominantly toroidal \citep{Guillochon&McCourt2017,Bonnerot+2017}. These toroidal magnetic fields modify the growth rates of the magnetorotational instability (MRI), which amplifies the toroidal fields. Eventually, the MRI growth rate can be suppressed by magnetic tension, leading to the saturation of the magnetic fields  \citep{Pessah&Psaltis2005,Begelman&Pringle2007}. Previous studies debated whether, in the absence of a relatively strong net vertical field, very strong toroidal magnetic flux could remain inside the disk or would escape vertically  \citep{JohansenLevin2008,Salvesen+2016}. The most recent work on this subject finds that strongly magnetized disks can self-sustain toroidal fields \citep{Squire2024}. 

Sufficiently strong magnetic fields can stabilize the disk against thermal and viscous perturbations, provided that magnetic pressure is powerful enough to compete with radiation pressure. The hypothesis that strong magnetic fields stabilize real accretion disks has been substantiated by global, self-consistent 3D radiation MHD simulations \citep{Sadowski2016,Jiang+2019,Huang+2023}, and more idealized studies \citep{Dexter&Begelman2019,Pan+2021,Sniegowska+2023}. Magnetic fields comparable in strength to those generated in the aforementioned analytic criteria and MHD simulations were shown by \citet{Kaur+2023} to be adequate for stabilizing TDE disks, but so far this has only been demonstrated for steady state disk solutions.

In this paper, we go further and investigate a 1D, time-dependent $\alpha$-disk model where magnetic fields dominate the pressure\footnote{We note that while we consider magnetic fields capable of dominating the total pressure budget, these ``magnetically elevated'' disks generally do not fall into the regime of magnetically arrested (i.e. ``MAD;'' \citealt{Narayan+2012}) accretion.}, aiming to study the late-time TDE evolution. Although late-time disks are dim and observationally challenging to study, they are significantly simpler to model compared to the highly complex \citep{Bonnerot&Stone2021} early-time TDE behavior. As a result, extracting astrophysical properties, such as the SMBH mass, from early-time behavior remains a considerable challenge. In principle, late-time disks should be simple to model, and therefore useful for estimating astrophysical parameters \citep{Mummery&vanVelzen2024}. However, achieving this goal requires first developing the most accurate model possible.

In \S\ref{sec: disk model}, we describe the basic equations of our model and solve them using both numerical methods and the self-similar solution. \S\ref{sec: linear viscosity} compares our model with the linear viscosity model. \S\ref{sec: observability} explores the detectability of TDEs decades or even centuries after their peaks. We conclude our findings in \S\ref{sec: conclusions}.

\section{Magnetized TDE Disks}\label{sec: disk model}

We develop a 1D, time-dependent thin disk model for a highly magnetized accretion disk, based on the standard Shakura-Sunyaev model \citep{Shakura&Sunyaev1973}. We use Newtonian gravity, as our main focus is on late-time TDEs when the relevant disk radii, $R$, are large and non-relativistic. The gas in the disk moves around the SMBH with a Keplerian angular frequency $\Omega_{\rm K}=\sqrt{GM_\bullet/R^3}$, where $M_\bullet$ is the SMBH's mass and $G$ is the gravitational constant. The matter in the disk is steadily accreted into the SMBH; the disk's surface density $\Sigma(t,R)$ evolves according to the disk diffusion equation: 
\begin{equation}
\frac{\partial\Sigma}{\partial t}=\frac{3}{R}\frac{\partial}{\partial R}\left\{R^{1/2}\frac{\partial}{\partial R}\left[\nu\Sigma R^{1/2}\right]\right\}    
\label{eq:diffusion eq}
\end{equation}
where $\nu$ is the effective (e.g. turbulent) viscosity. We follow the $\alpha$-disk prescription $\nu=\alpha H c_{\rm s}$ \citep{Shakura&Sunyaev1973}. The disk height $H=c_{\rm s}/\Omega_{\rm K}$ is determined from the vertical hydrostatic equilibrium, where $c_{\rm s}=\sqrt{P/\rho}$ is the speed of sound. The density inside the disk is $\rho=\Sigma/H$, and $P$ is the disk pressure. 

\begin{figure}
    \includegraphics[width=85mm]{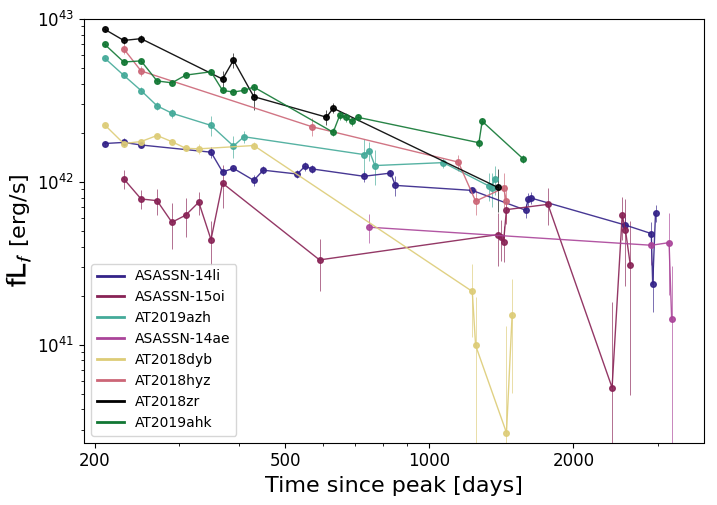}
\caption{Late-time UV light curve observations from \textit{Swift}/UVOT for various TDEs \citep{Mummery+2024}. Data are presented for the uvw2 filter. The light curves have been binned and averaged over $20$-day intervals, with error bars representing (asymmetric) 1$\sigma$ uncertainties. The observations generally show smoothly evolving late-time emission without the dramatic instabilities predicted by unmagnetized $\alpha$-disk models (\citealt{Shen&Matzner2014, Piro&Mockler2024}, with the possible exception of AT2018dyb).
}
\label{fig: late-time observations}
\end{figure}

When the pressure in the disk is radiation-dominated, \cref{eq:diffusion eq} describes an unstable disk \citep{Piran1978}, where the surface density and temperature cycle between high and low accretion states \citep{Shen&Matzner2014,Piro&Mockler2024}. However, as shown in \cref{fig: late-time observations}, current late-time TDE observations do not generally display evidence of such instabilities in their late-time light curves, with one possible exception (AT2018dyb; \citealt{Leloudas+2019}). Therefore, in this paper, we \blue{follow \citet{Kaur+2023} and} focus on magnetized accretion disks, assuming that the disk pressure, $P$, is dominated by magnetic fields, with $P\approx P_{\rm m}=B^2/8\pi$, where $B$ is the magnetic field strength.

To estimate the magnetic field strength, we assume a saturation criterion for the MRI produced when toroidal magnetic fields become ``super-thermal'' \citep{Pessah&Psaltis2005, Begelman&Pringle2007}. While other saturation criteria have been proposed as well \citep{Oda+2009,Begelman&Armitage2023}, the one we employ has been roughly validated in recent MHD simulations \citep{Jiang+2019,Mishra+2022,Huang+2023}. The MRI is suppressed when the Alfv\'en speed $v_{\rm A}=\sqrt{P_{\rm m}/\rho}$ exceeds the geometric mean of the Keplerian velocity $v_{\rm K}=\sqrt{GM_\bullet/R}$, and the gas sound speed $c_{\rm s,g}=\sqrt{P_{\rm g}/\rho}$, where $P_{\rm g}=\rho\frac{k_{\rm B}T}{\mu m_{\rm p}}$ is the gas pressure. The resulting magnetic pressure is given by: 
\begin{equation}
    P_{\rm m}=v_{\rm K}\rho\sqrt{\frac{k_{\rm B}T}{\mu m_{\rm p}}}
    \label{eq: mag pressure}
\end{equation}
where $k_{\rm B}$ is the Boltzmann constant, $\mu=0.6$ is the mean molecular weight for Solar metallicity gas, $m_{\rm p}$ is the proton mass, and $T$ is the mid-plane temperature of the disk.

\blue{At the high, trans- or super-Eddington accretion rates typical of early-time TDE disks, the \citet{Begelman&Pringle2007} ansatz above does not guarantee that $P_{\rm m} \approx P$, as radiation pressure will continue to dominate unless $P_{\rm m}$ is even larger than the value in Eq. \ref{eq: mag pressure}.  However, at the low Eddington ratios relevant for the late-time TDE disks we aim to investigate, Eq. \ref{eq: mag pressure} is sufficient to guarantee magnetic stabilization of the disk (and dominance of $P_{\rm m}$ over radiation pressure) at essentially all relevant radii, see e.g. \citet{Kaur+2023}, Fig. 1.}

We note that it is unknown whether angular momentum transport in magnetized disks is primarily driven by local viscous torques, as our use of the $\alpha$-ansatz assumes. Previous research has also examined situations where, instead, large-scale magnetic torques may dominate the angular momentum loss in magnetized disks   \citep{Blandford&Payne1982,Ferreira&Pelletier1993,Ferreira&Pelletier1995,Begelman2024, Tamilan+2024}. Still, these disks may not appear significantly different from standard disks \citep{Begelman2024}. 

The temperature of the disk is given by the energy equation. We assume a balance between viscous heating and radiative cooling:
\begin{equation}
    \frac{4\sigma_{\rm SB}T^4}{3\kappa_{\rm{es}}\Sigma}=\frac{9}{8}\nu\Sigma\frac{GM_\bullet}{R^3}
\end{equation}
where $\sigma_{\rm SB}$ is the Stefan-Boltzmann constant and $\kappa_{\rm{es}}=0.34\,\rm{cm}^2\rm{g}^{-1}$ is the Thomson scattering opacity. Electron scattering opacity dominates in the inner regions that interest us and therefore we neglect absorption. Additionally, we neglect advective cooling since, at late times, the disks are thin and sub-Eddington (this neglect would not be justified at early times; conversely, however, advective early-time TDE disks would likely be stable even absent strong magnetic fields; \citealt{Kaur+2023}).   

We assume the disk is optically thick, and the emission is isotropic, so the spectral luminosity \red{$L_\nu$}\blue{$L_f$} at a frequency \red{$\nu$}\blue{$f$} is given by Planck's blackbody distribution, \red{$B_\nu$}\blue{$B_f$}, as described in
\begin{equation}
    \red{L_\nu}\blue{L_f}=4\pi^2 \int_{R_{\rm in}}^{R_{\rm out}} \red{B_\nu}\blue{B_f}\left(T_{\rm eff}\right)R dR,
\label{eq: light curves}
\end{equation}
 where $T_{\rm eff}=T\left(\frac{4}{3\kappa_{\rm es}\Sigma}\right)^{1/4}$ is the effective temperature\footnote{For simplicity we ignore the color corrections to emission that cause quasi-thermal accretion disk spectra to deviate from perfect blackbodies \citep{ShimuraTakahara1995}.  While these corrections can cause $\sim 50\%$ differences between color temperature and effective temperature in X-ray bands, the effect is much smaller at longer wavelengths due to the increased absorption opacity in cooler parts of the disk \citep{DavisElAbd2019}.}. 

The effective viscosity is an approximation to the transport of angular momentum by magnetized turbulence (driven by the MRI)\blue{. Under the assumption of an $\alpha$-disk prescription, it is given by $\nu=\alpha Hc_{\rm s}=\frac{\alpha c_s^2}{\Omega_{\rm K}}=\frac{\alpha P}{\Omega_{\rm K}\rho}$}, and when the pressure in the disk is dominated by magnetic fields \blue{(\cref{eq: mag pressure})}, it becomes equal to: 
\begin{equation}
\begin{aligned}
    \nu_{\rm mag}&=\nu_{0,\rm{m}}R^{5/7}\Sigma^{2/7}\\
    \nu_{0,\rm{m}}&=\left[\frac{27}{32\sigma_{\rm SB}}\alpha^8\kappa_{\rm{es}}GM_\bullet\left(\frac{k_{\rm B}}{\mu m_{\rm p}}\right)^{\blue{4}}\right]^{1/7}.
\end{aligned}
\label{eq:viscosity}
\end{equation}
Because $\nu$ is a function of both $R$ and $\Sigma$ the evolution equation \cref{eq:diffusion eq} becomes a nonlinear diffusion equation. This equation can be separated into a trivial time equation and a nonlinear ordinary differential equation for the radius (see Appendix \ref{appendix: separability Emden-Fowler}, where we show for the first time that a broad class of $\alpha$-viscosity laws can reduce the disk diffusion partial differential equation to the Emden-Fowler ordinary differential equation). However, there is currently  no general analytical solution for this equation, necessitating the use of numerical methods.

\subsection{Numerical Solution}
In this paper, we integrate \cref{eq:diffusion eq} numerically. We set the inner boundary to be the innermost stable circular orbit (ISCO), which for a non-spinning BH equals to $R_{\rm in}=6r_{\rm g}$, where $r_{\rm g}=GM_\bullet/c^2$ is the gravitational radius of the central BH and $c$ is the speed of light. We assume $\Sigma=0$ at the inner boundary, implies that the viscous torque there is zero, meaning that all the matter that reaches the ISCO is accreted along with its angular momentum. The outer boundary is always chosen to be $R_{\rm out}\gg R_{\rm in}$ such that the surface density $\Sigma$ vanishes at some radius smaller than $R_{\rm out}$ at all times.  

In our fiducial models, we use a model in which, initially, half of the mass from the disrupted star has escaped to infinity, while the other half has already circularized to form a ring-like disk at \citep{Strubbe+2009}:
\begin{equation}
    r_{\rm c}=\frac{1+e_\star}{\beta}r_{\rm t}=2r_{\rm t}
\end{equation}
where $e_\star$ is the orbital eccentricity of the initially approaching star, and $\beta$ is the ratio of the tidal disruption radius $r_{\rm t}$ to the pericenter distance, and in the final equality we chose $e_\star=1$ and $\beta=1$. 
The tidal radius is \citep{Rees1988}: 
\begin{equation}
\begin{aligned}
    r_{\rm t}&=\left(\frac{M_\bullet}{M_\star}\right)^{1/3}R_\star\\
    &\sim 16\left(\frac{M_\bullet}{10^6M_\odot}\right)^{-2/3}\left(\frac{M_\star}{0.1M_\odot}\right)^{-1/3}\left(\frac{R_\star}{0.16R_\odot}\right) r_{\rm g}
\end{aligned}
\end{equation}
where $M_\star$ and $R_\star$ are the mass and radius of the disrupted star, respectively. We assume a lower-main-sequence star such that $R_\star=(M_\star/M_\odot)^{0.8}R_\odot$ \citep{Kippenhahn+2012}. 

Initially, the disk has a Gaussian distribution centered around $r_{\rm c}$:
\begin{equation}
    \Sigma(R,t=0)=\frac{M_\star}{4\pi^{3/2}r_{\rm c}\sigma}\exp\left[-\frac{(R-r_{\rm c})^2}{\sigma^2}\right]+\epsilon
\end{equation}
where $\sigma=0.1r_{\rm c}$, and $\epsilon$ is a small density floor that is many orders of magnitude lower  than the peak of the initial Gaussian. We note that the prefactor in the Gaussian is determined by setting the initial total mass contained within the ring to $M_\star/2$.

As a second scenario, instead of starting with a Gaussian initial condition, we also explore \cref{eq:diffusion eq} with a source term that reflect the slow accumulation of the bound debris into the accretion disk. Initially, the surface density is a small value, and over time mass is added to the disk around $r_{\rm c}$ with a Gaussian distribution. The late-time rate of mass fallback onto the disk is \citep{Stone+2013}:
\begin{equation}
    \dot{M}_{\rm fall}=\frac{M_\star}{3t_{\rm fall}}\left(\frac{t+t_{\rm fall}}{t_{\rm fall}}\right)^{-5/3}, \label{eq:fallback}
\end{equation}
where
\begin{equation}
    t_{\rm fall}=3.5\times 10^6 \rm{sec}\left(\frac{M_\bullet}{10^6M_\odot}\right)^{1/2}\left(\frac{M_\star}{M_\odot}\right)^{-1}\left(\frac{R_\star}{R_\odot}\right)^{3/2}.
\end{equation}

\begin{figure}
    \includegraphics[width=85mm]{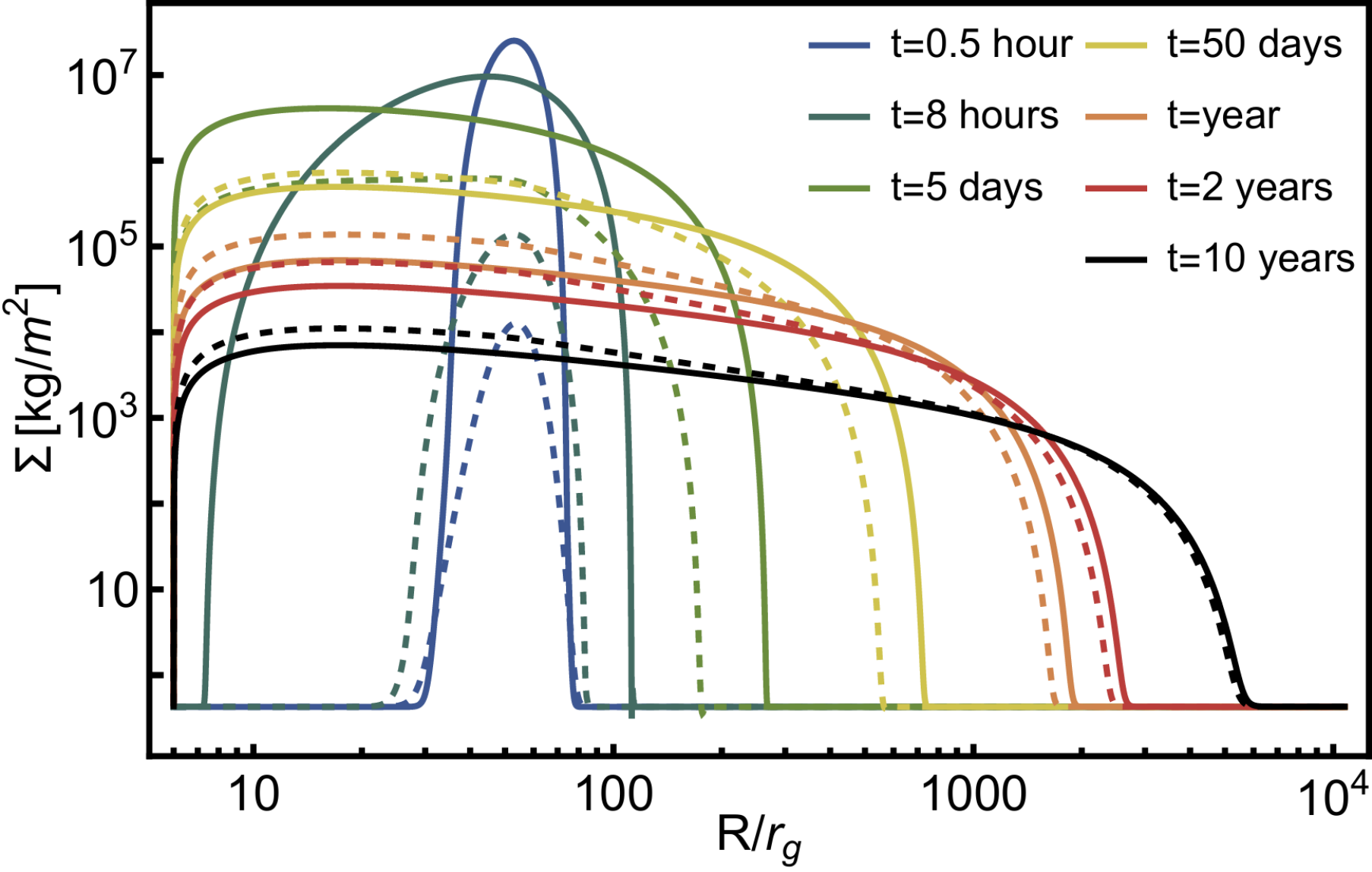}\par 
    \includegraphics[width=85mm]{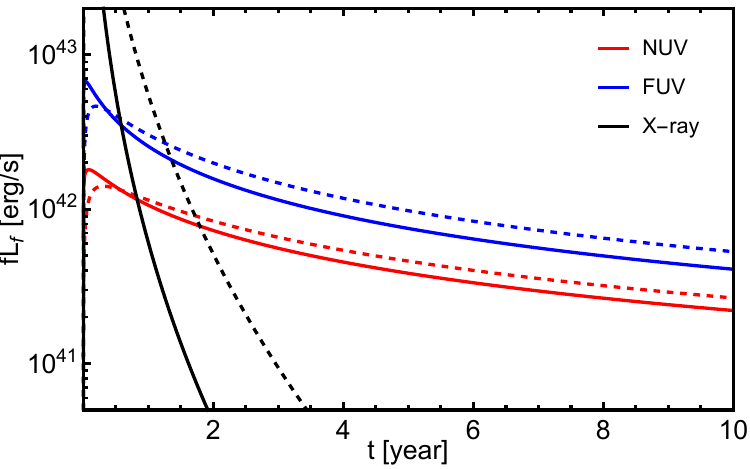}\par
\caption{A comparison between the Gaussian initial condition (solid) to the source function (dashed). The top panel shows the surface density as a function of radius at different times, while the bottom panel presents the light curves at different wavelengths (NUV: $\lambda=250 \text{nm}$, FUV: $\lambda=150 \text{nm}$, and X-ray: $\red{h\nu}\blue{hf}=300 \text{eV}$). The masses of the SMBH and the disrupted star are $M_\bullet=10^6M_\odot$ and $M_\star=0.3M_\odot$ respectively, and $\alpha=0.1$.}
\label{fig: Gaussian vs source function}
\end{figure}

The source function method is probably a more realistic solution than the initial Gaussian distribution at early times, although both approaches gloss over the considerable complexity \citep{Hayasaki+2013, Shiokawa+2015, Bonnerot&Stone2021} of the circularization process. However, in this paper, we are interested in the late stages of the TDE, typically years after disruption. After such a long period, the initial conditions in our 1D models are effectively forgotten \citep{Cannizzo+1990}, and there is not a big difference between the two prescriptions for mass injection. In \cref{fig: Gaussian vs source function}, we compare the two methods, showing the surface density as a function of radius at different times, as well as the light curves. Although the surface densities differ quite a bit initially, they converge to the same solution over time. The UV light curves are also similar at late times, and while the X-ray luminosity {\it is} different, it decays rapidly and becomes effectively undetectable \citep{Lodato&Rossi2011}. Therefore, for simplicity, we use only the Gaussian initial condition in this paper unless specified otherwise.

\subsection{Self-Similar solution}
Although the evolution equation is a nonlinear equation without an exact analytic solution, in the case of highly magnetized accretion disks, where the viscosity\footnote{Self-similar solutions can also exist for other types of strongly magnetized disk models, e.g. those where angular momentum transport is dominated by wind losses \citep{Tamilan+2025}.} follows a power law function of $\Sigma$ and $R$ (see  \cref{eq:viscosity}), a self-similar solution for $\Sigma$ does exist \citep{Lin&Pringle1987,Cannizzo+1990, Pringle1991}. In this self-similar solution, the inner boundary is at $R_{\rm{in}}=0$, and there are two types of solutions: (i) a constant disk mass with increasingly disk angular momentum, and (ii) a disk with zero torque at the origin. In this paper, we consider the latter solution, and for our model of a magnetized accretion disk the similarity solution is given by:
\begin{equation}
\begin{aligned}
    \Sigma_{\rm m,S}(t,R)&=\Sigma_{0,\rm S}\left(\frac{R}{R_{0,\rm S}}\right)^{-5/9}\left(\frac{3\nu_{0,\rm S}t}{4R_{0,\rm S}^2}\right)^{-35/36}\times\\
    &\times\left[1-\frac{1}{52}\left(\frac{R}{R_{0,\rm S}}\right)^{13/9}\left(\frac{3\nu_{0,\rm S}t}{4R_{0,\rm S}^2}\right)^{-13/18}\right]^{7/2}
\end{aligned}
\label{eq: magnetic self-similar solution}
\end{equation}
where $R_{0,\rm S}$, $\Sigma_{0,\rm S}$ and $\nu_{0,\rm S}$ are arbitrary constants. 

\begin{figure}
\includegraphics[width=85mm]{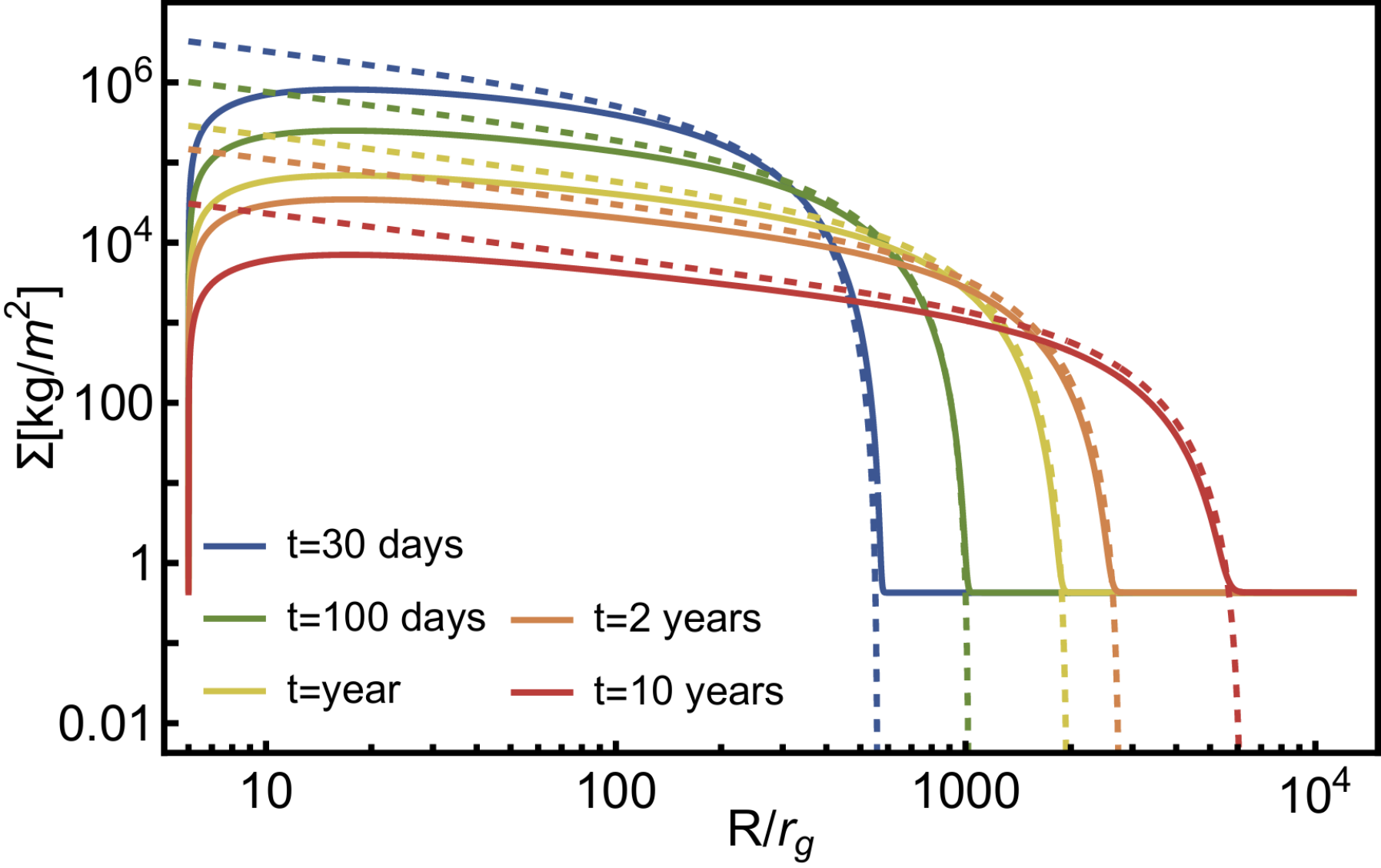}
\caption{Surface density $\Sigma$ plotted against dimensionless radius $R/r_{\rm g}$ at late times. The solid curves are the numerical solution of the disk diffusion equation while the dashed curves are the self-similar solution (Eq. \ref{eq:similaritysol}). The masses of the SMBH and the disrupted star are $M_\bullet=10^6M_\odot$ and $M_\star=0.3M_\odot$ respectively, and $\alpha=0.1$.  The self-similar solutions provide a good match to numerical integration of the disk diffusion equation at late times and large radii.}
\label{fig:self-similar solution}
\end{figure}

The self-similar solution is not the exact solution to \cref{eq:diffusion eq}, but it closely approximates the behavior of the system at late times. In \cref{fig:self-similar solution}, we compare the self-similar solution to the numerical solution for $M_\star=0.3M_\odot$. For the comparison, we need to determine $R_{0,\rm S}$, $\Sigma_{0,\rm S}$ and $\nu_{0,\rm S}$. By choosing $\nu_{0,\rm S}=\nu_{\rm mag}(R_{0,\rm S},\Sigma_{0,\rm S})$, we can rewrite \cref{eq: magnetic self-similar solution} with only one free parameter: 
\begin{equation}
\begin{aligned}
    \Sigma_{\rm m,S}(t,R)&=\left(\frac{\nu_{0,\rm S}}{\nu_{0,\rm{m}}}\right)^{7/2}R^{-5/9}\left(\frac{3\nu_{0,\rm S}t}{4}\right)^{-35/36}\times\\
    &\times\left[1-\frac{1}{52}R^{13/9}\left(\frac{3\nu_{0,\rm S}t}{4}\right)^{-13/18}\right]^{7/2} 
\end{aligned}
\label{eq:similaritysol}
\end{equation}
where $\nu_{0,\rm{m}}$ is calculated at \cref{eq:viscosity}. \blue{The outer radius of the disk in the self-similar solution is $R_{\rm out}=52^{9/13}\left(\frac{3\nu_{0,\rm S}t}{4}\right)^{1/2}\propto t^{1/2}$.} To determine the remaining free parameter, $\nu_{0,\rm S}$, we calculated the outer radius of the numerical disk, i.e. the radius where the surface density drops to the numerical floor $\epsilon$, as a function of time (between $1$-$10$ years). We then fixed $\nu_{0,\rm S}$ to best fit the outer radius at these times. \cref{fig:self-similar solution} shows that the self-similar solution closely resemble the numerical solution, with some differences: at large radii, the numerical solution decays to the numerical floor $\epsilon$, while the self-similar solution drops to zero. Additionally, differences in the inner boundary conditions result in different surface density profiles at small radii. In Appendix \ref{appendix: self-similar fit}, we provide a fitting function for $\nu_{0,\rm S}$, and present the relative errors between the light curves obtained from the numerical and the self-similar solutions.

\section{Comparison to Linear Viscosity}\label{sec: linear viscosity}

In time-dependent 1D accretion disk models where radiation pressure dominates, the disk exhibits viscous and thermal instabilities. However, as described in Sec. \ref{sec: intro}, observations of many such astrophysical systems reveal disks that remain stable. A common theoretical solution is to modify the viscosity parametrization to a linear form \citep[e.g.][]{Mummery2019}: 
\begin{equation}
    \nu_{\rm l}=\nu_{0,\rm {l}}R^{\mu_{\rm l}}.
\end{equation}

This linear viscosity approach is popular because it stabilizes the disk and is also theoretically simple, offering an analytical solution for \cref{eq:diffusion eq}, see for example \citet{Metzger+2012}.  However, this choice of effective viscosity is generally made for numerical convenience and lacks a solid physical basis. Therefore, we aim to compare 1D disks evolving under linear viscosity model with strongly magnetized $\alpha$-disks. To do this, we explore various values of the power $\mu_{\rm l}$, and different methods for determining the normalization factor $\nu_{0,\rm {l}}$.  

\blue{One value of $\mu_{\rm l}$ that may be relevant at late times can be derived from the self-similar solution. From \cref{eq: magnetic self-similar solution}, we find that $\Sigma\propto R^{-5/9}$, and substituting this into \cref{eq:viscosity} gives $\nu_{\rm mag}\propto R^{\mu_{\rm l}}$, where $\mu_{\rm l}=5/9$. Throughout this section, we also explore other values for the power law index: $\mu_{\rm l}=\{3/2,0,-3/2\}$.}

\begin{figure}
\includegraphics[width=85mm]{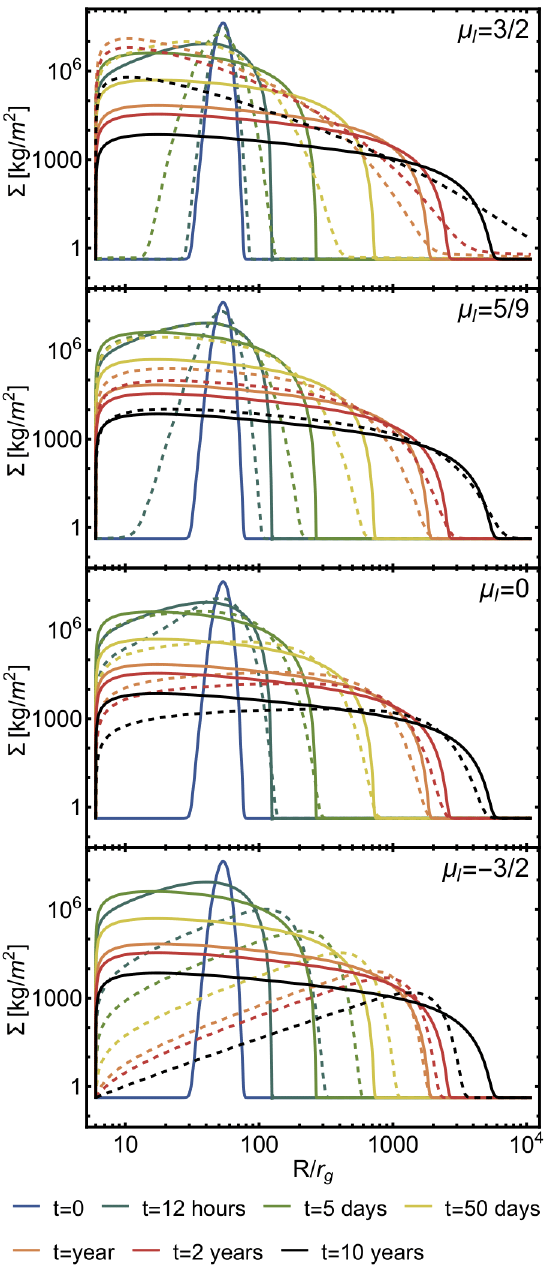}
\caption{Surface densities plotted against radius at different times, for the magnetized viscosity (solid) and the linear viscosity (dashed). The power laws of the linear viscosity are $\mu_{\rm l}=\{3/2\blue{,5/9},0,-3/2\}$ from the top panel to the bottom panel. The normalization factor of the linear parametrization $\nu_{0,\rm l}$ is determined by equating the outer radius of the two disks after 1 year. The choice of $M_\bullet$, $M_\star$, and $\alpha$ is the same as in \cref{fig:self-similar solution}. \red{The best match between the surface densities occurs for $\mu_{\rm l}=0$.}\blue{The best match between the surface densities occurs for $\mu_{\rm l}=5/9$ in the inner regions of the disk and $\mu_{\rm l}=0$ in the outer regions.} }
\label{fig:linear-mag comparison surface density}
\end{figure}

In \cref{fig:linear-mag comparison surface density}, we compare the surface densities $\Sigma$ between the linear viscosity and the magnetized disk models \red{for $\mu_{\rm l}=\{3/2,0,-3/2\}$} at different times. The initial and boundary conditions are identical across all cases, and because we are particularly interested in late times, we determined the normalization factor $\nu_{0,\rm l}$ by equating the outer radii of the disks at 1 year (we explore different choices in Appendix \ref{appendix: more linear viscosity comparison}). In both the linear and the magnetized viscosity cases, the disk is stable, and the material that was once concentrated in a ring will gradually spread. Additionally, the surface densities at the outer radii are similar due to our choice of the normalization factor, but at small radii they \red{are quite different}\blue{can differ significantly}. \red{The best match between the surface density profiles occurs for $\mu_{\rm l}=0$, with worse agreement for other values of $\mu_{\rm l}$.}\blue{The best match between the surface density profiles occurs for $\mu_{\rm l}=5/9$ at late-times. However, $\mu_{\rm l}=0$ provides the best agreement with the time evolution in the outer regions of the disk. Other values of $\mu_{\rm l}$ show worse agreement with the behavior of the magnetized disk.} 

\begin{figure}
\includegraphics[width=85mm]{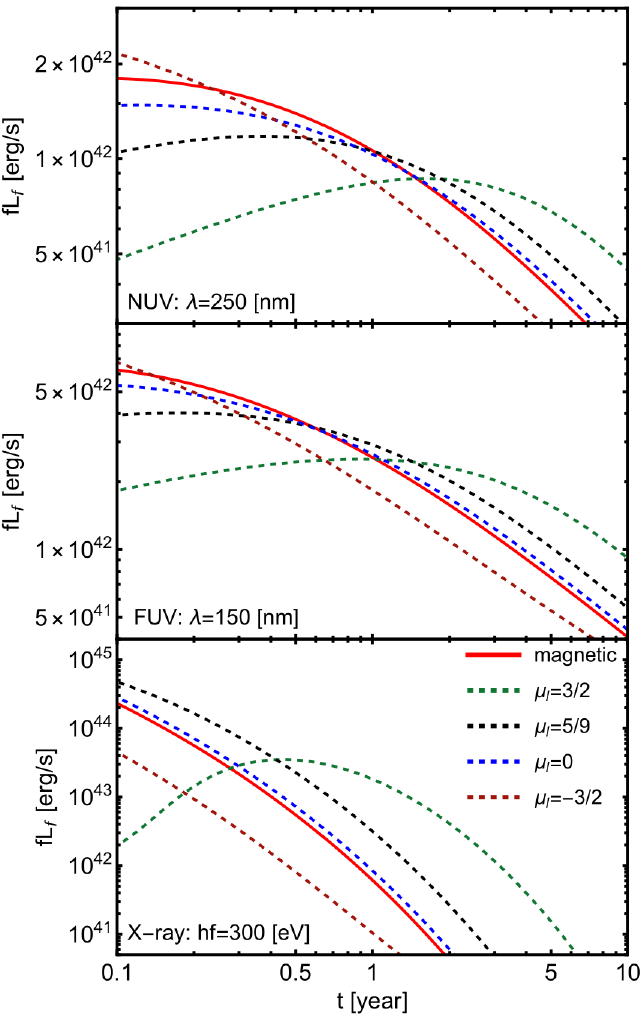}
\caption{Light curves for near-UV, far-UV and X-ray bands (top panel to bottom panel) with Gaussian initial conditions for magnetized viscosity solutions (solid red) and linear viscosity solutions (dashed), with $\mu_{\rm l}=\{3/2\blue{,5/9},0,-3/2\}$ (\blue{green}, black, blue, brown). The normalization factor of the linear parametrization $\nu_{0,\rm l}$ is determined by equating the outer radius of the disk after 1 year. The choice of $M_\bullet$, $M_\star$, and $\alpha$ is the same as in \cref{fig:self-similar solution}. The light curves of the linear and magnetized viscosity models appear very similar for $\mu_{\rm l}=0$, but quite distinct for other values of $\mu_{\rm l}$.}
\label{fig:linear-mag comparison light curve}
\end{figure}

However, we cannot observe the surface density directly, so we also need to look at the light curves. Therefore, in \cref{fig:linear-mag comparison light curve} we compare the light curves at three different wavelengths (near-UV, far-UV and X-ray), for both linear and $\alpha$-viscosity models. Surprisingly, the light curves for magnetic viscosity and linear viscosity with $\mu_{\rm l}=0$ looks very similar, even for the X-ray band that originates from the inner radii of the disk \citep{Lodato&Rossi2011}; other choices of $\mu_{\rm l}$ produce much worse agreement.

To understand the similarity of the light curves between the magnetized $\alpha$-disk and the linear viscosity solutions with $\mu_{\rm l}=0$, we can check the time evolution of the self-similar solutions. From \cref{eq: magnetic self-similar solution} the time evolution of the surface density of the magnetized disk is $\Sigma_{\rm m,S}\propto t^{-35/36}$, while the time evolution of the linear viscosity disk is $\Sigma_{\rm l,S}\propto t^{-(5-2\mu_{\rm l})/(4-2\mu_{\rm l})}$ \citep{Metzger+2008}. For $\mu_{\rm l}=0$, the power laws are ultimately different, as can be seen in \cref{fig:linear-mag comparison surface density}. However, in steady state, the luminosity $L\propto\dot{M}\propto\nu\Sigma$ where $\dot{M}$ is the mass accretion rate. When we substitute the self-similar solutions into the luminosity, we find:
\begin{equation}
    \begin{aligned}
        L_{\rm m,S}&\propto t^{-5/4}\\
        L_{\rm l,S}&\propto t^{-(5-2\mu_{\rm l})/(4-2\mu_{\rm l})}
    \end{aligned}
\end{equation}
which for $\mu_{\rm l}=0$ reduce to precisely the same power-laws. This means that analytic Green's function solutions with $\mu_{\rm l}=0$ can be used as a decent approximation to a magnetized accretion disk with $\alpha$-viscosity; these solutions take the form of Bessel functions in Newtonian gravity \citep[e.g.][]{Metzger+2008} or hypergeometric functions in relativistic gravity \citep{Mummery2023}. 

\section{Observability of Late-Time TDE Disk Emission }\label{sec: observability}

\begin{figure*}
\centering
\includegraphics[width=85mm]{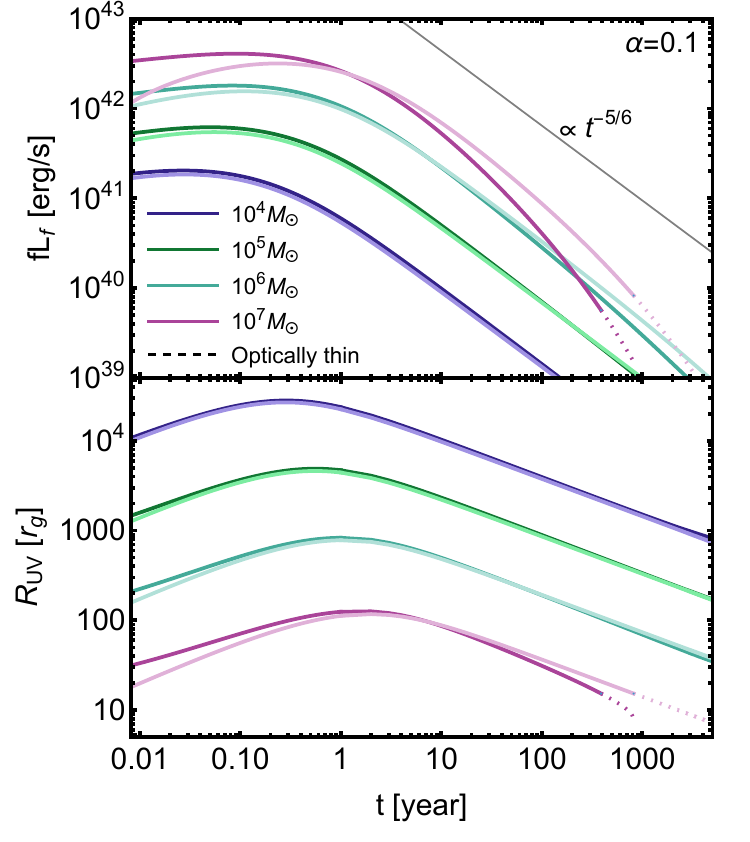}
\includegraphics[width=85mm]{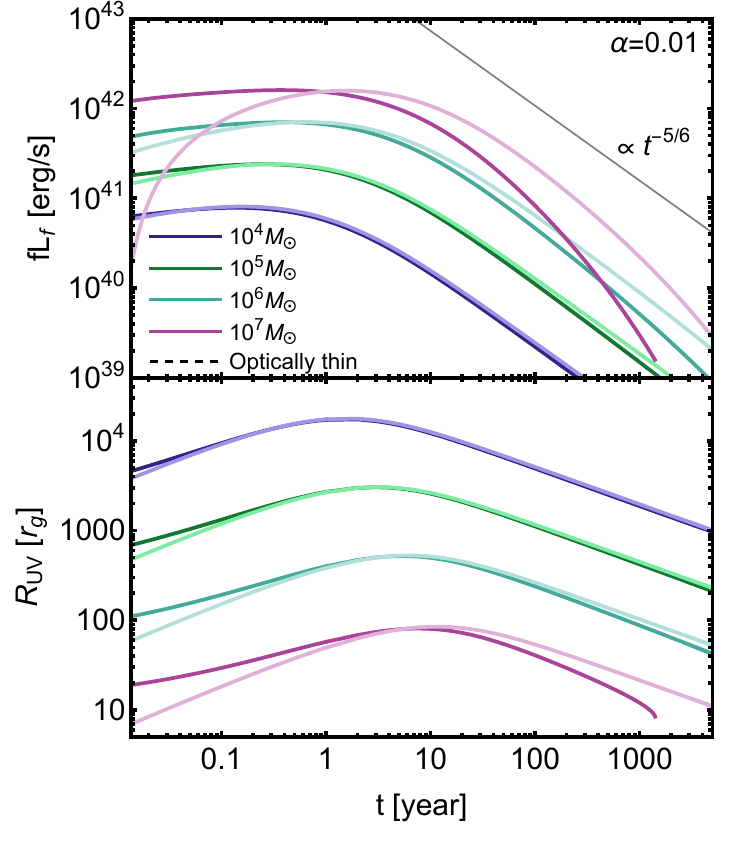}
\caption{Light curves (top panels) and $R_{\rm{UV}}$ (bottom panels) for the magnetized $\alpha$-disk, plotted as functions of time, and calculated for various $M_\bullet$ values at a NUV wavelength  ($\lambda=250\rm{nm}$). The left panels correspond to $\alpha = 0.1$, while the right panels use $\alpha = 0.01$. Numerical solutions are shown in darker colors, while self-similar solutions are represented with lighter colors (the relative errors between the numerical and the self-similar solutions are calculated in Appendix \ref{appendix: self-similar fit}). The curves transition to dashed lines when the disks become optically thin and are truncated when $R_{\rm{UV}}$ shrinks to the ISCO. A gray reference line, scaled as $L_{\rm m,UV}\propto t^{-5/6}$ (Eq. \ref{eq:scaling}) is included for comparison.  The light curves remain UV-bright for many years before eventually turning off. }
\label{fig: light curves- RUV}
\end{figure*}

Another insight from \cref{fig:linear-mag comparison light curve} is that TDEs remain UV-bright for many years. In this section, we aim to determine for just how long a TDE remains observable. 
 \cref{fig: light curves- RUV} shows light curves at a UV wavelength of $\lambda=250 \rm{nm}$ over thousands of years for different SMBH masses, and the Shakura-Sunyaev $\alpha$ parameters. As the SMBH mass increases, the TDE is brighter \citep{Mummery+2024}, and for larger values of $\alpha$, the light curves evolve faster. To study the UV light curve, we define $R_{\rm UV}$ as the disk annulus where the temperature corresponds to the UV wavelength: $T_{\rm eff}(R_{\rm UV})=T_{\rm UV}=1.2\times10^4\left(\frac{\lambda}{250 \rm{nm}}\right)\rm{K}$. At this radius, the luminosity is $L_{\rm UV}\propto R^2_{\rm UV}\sigma_{\rm SB}T^4_{\rm UV}\propto \left(\nu\Sigma\right)^{2/3}$. Substituting the self-similar solution for a magnetized disk, the UV luminosity evolves over time as: 
\begin{equation}
    L_{\rm m,UV}\propto t^{-5/6}. 
    \label{eq:scaling}
\end{equation}

Eventually, however, the TDE disk will turn off in the UV, either when the $R_{\rm UV}$ annulus becomes effectively optically thin or when it shrinks to the ISCO (in reality there will still be some UV emission after one of these criteria is satisfied, but it will become dimmer much more rapidly than in Eq. \ref{eq:scaling}). The disk can be treated as a blackbody when the effective optical depth $\tau_{\rm eff}=\sqrt{3\kappa_{\rm abs}(\kappa_{\rm es}+\kappa_{\rm abs})}\Sigma>1$ for a given disk annulus, where $\kappa_{\rm abs}$ is the absorption opacity, approximated here by Kramers' opacity $\kappa_{\rm abs}\sim \kappa_{\rm Kr}=1\times10^{24}\left(\frac{\rho}{\rm g/cm^3}\right) \left(\frac{T}{\rm K}\right)^{-7/2}\rm{cm^2\,g^{-1}}$. As shown in \cref{fig: light curves- RUV}, at larger SMBH masses, the disk will become optically thin more quickly, and $R_{\rm UV}$ will reach the ISCO sooner than is the case for smaller $M_\bullet$ values.  Sometimes, the UV-emitting regions will become optically thin (in the effective opacity sense) before reaching the ISCO, the alternate ordering of events is also possible. Late-time TDEs are more likely to turn off in the UV due to optical depth effects when $\alpha$ is large and more likely to turn off due to $R_{\rm UV} \to R_{\rm ISCO}$ when $\alpha$ is small.

\subsection{ASASSN–14li}

ASASSN-14li is a TDE that discovered by the All-Sky Automated Survey for SuperNovae (ASAS-SN; \citealt{Shappee+2014, Holoien+2016}). It was first detected on 2014 November 22, in a post-starburst galaxy at a redshift of $z=0.0206$ \citep{Jose+2014}. The late-time optical and UV light curves are well observed \citep{Brown+2017}, showing gradual evolution that is consistent with emission from an accretion disk \citep{vanVelzen+2019}.

Since ASASSN-14li has been observed over much of the past decade, it serves as an ideal candidate for late-time light curve modeling. In this subsection, we analyze its late-time light curves\footnote{Data taken from \citet{Mummery+2024}.} (starting 380 days after its disruption) obtained with \textit{Swift}/UVOT across five filters: B, U, UVM2, UVW1, and UVW2. Using the magnetized $\alpha$-viscosity model, we fit the data to estimate the Shakura-Sunyaev $\alpha$ parameter. For this analysis, we adopt a SMBH mass of $\log_{10}(M_\bullet/M_\odot)=6.23^{+0.39}_{-0.40}$, based on velocity dispersion\footnote{This mass estimate from galaxy scaling relations is statistically consistent with X-ray continuum fitting mass estimates \citep{Wen+2020}, though the error bars on both are significant.} measurements \citep{Wevers+2017}. 

We fit the observed light curves using those calculated from \cref{eq: light curves} with the self-similar solution. The free parameters in our model are the Shakura-Sunyaev $\alpha$ parameter and $\nu_{0,\rm S}$, which corresponds to the stellar mass. The total likelihood is computed as follows: 
\begin{equation}
    \mathcal{L}=-\sum\limits_{\text{band},i}\sum\limits_{\text{data},j} \frac{\left(O_{i,j}-M_{i,j}\right)^2}{E_{i,j}^2},
\end{equation}
where $O_{i,j}$, $M_{i,j}$ and $E_{i,j}$ represent the observed light curves, the model light curves, and the uncertainty of the $j$-th data point in the $i$-th band, respectively. 

The best fit parameters obtained are $\alpha=0.0170\pm 0.0001$ and $\nu_{0,\rm S}=1.4\times10^{14}$, which corresponds to a stellar mass of $M_\star=0.12M_\odot$.
The error in $\alpha$ represents only the statistical uncertainty. Additionally, there are systematic errors arising from the SMBH mass and uncertainties in our model, which we do not quantify in this preliminary exploration. The errors due to the use of the self-similar solution are calculated in Appendix \ref{appendix: self-similar fit}.

ASASSN-14li's viscous spreading is known to be slow \citep{Wen+2023}; however, the fitted parameter $\alpha$ falls within a theoretically reasonable range \citep{Hirose+2006}.  There is therefore no need to invoke additional physics to further limit the spread of the disk outer edge, such as angular momentum loss in a magnetized wind \citep{Wen+2023}.

\subsection{Fossil TDEs}

Even though the accretion disks of TDEs will eventually turn off, TDEs may remain bright in the UV for decades or centuries, allowing us to potentially detect fossil TDEs that erupted long ago. By fitting theoretical models to observations, we could measure parameters of the SMBHs, such as their mass, and thereby gain insights into their origins \citep{Volonteri2010} and growth histories \citep{Bhowmick+2024}, which often remain unclear. To accurately measure these parameters, however, we need to use the most precise models possible. Late-time TDE disks are much simpler phenomena than the early stages of a TDE, before an accretion disk has circularized, making it easier to develop theoretical models to fit observations. Therefore, it may be useful to build a large library of fossil TDEs. 

The Ultraviolet Transient Astronomy Satellite (\textit{ULTRASAT}), scheduled to be launched in 2027 \citep{Shvartzvald+2024}, is a promising instrument for detecting these events. \textit{ULTRASAT} is a near-UV space telescope ($230-290$nm), designed to conduct a wide-field UV survey for transients. The predicted TDE detection rate for \textit{ULTRASAT} is enormous and will expand the current TDE sample by $1-2$ orders of magnitude \citep{Shvartzvald+2024}, similar to the discovery potential of Rubin Observatory's LSST but without the false positives that are likely to trouble a lower cadence optical survey \citep{Bricman&Gomboc2020}. Given this potential, we want to check whether \textit{ULTRASAT} could also detect fossil TDEs. In this subsection, we focus on UV detection with \textit{ULTRASAT} rather than optical surveys like LSST because fossils are brighter in the UV than in optical wavelengths. Furthermore, galaxies are much brighter in optical wavelengths, meaning that late-time fossils would likely be overwhelmed by background starlight, making them effectively undetectable in the optical.

Detecting fossil TDEs presents different challenges compared to detecting the earlier stages of transients. While transient astronomy typically involves searching for changes in source brightness over a short time period, fossil TDEs exhibit much slower changes from a fainter baseline luminosity. To detect fossil TDEs, we need two things: (i) the source must be bright enough to be detectable in the first place, and (ii) it must evolve quickly enough to be classified as a transient rather than as a steady source. Therefore, we will need to monitor the source for an extended period, such as a year, and look for changes in luminosity over this time.

For detection, we assume a typical limiting AB magnitude of $m_{\rm AB}=22.5$ \citep{Shvartzvald+2024}. Since TDEs can be  observed in the UV up to cosmological distances, we account for K-corrections to determine the limiting luminosity distance for detecting fossil TDEs: 
\begin{equation}
    D_{\rm L}^{\rm max}(z,t,M_\bullet,M_\star)=\sqrt{\frac{(1+z)L_{(1+z)\red{\nu}\blue{f}}(t,M_\bullet,M_\star)}{4\pi F\red{_\nu}\blue{_f}}}
    \label{eq: luminosity distance}
\end{equation}
where \red{$L_\nu$}\blue{$L_f$} is the luminosity calculated from our model at the center of the \textit{ULTRASAT} band ($250$nm), $\red{F_\nu}\blue{F_f}=10^{-\frac{m_{\rm AB+48.6}}{2.5}}\approx 3.6\times10^{-29}\rm{\,erg\,s}^{-1}\,\rm{cm}^{-2}\,\rm{Hz}^{-1}$ is the limiting flux density of \textit{ULTRASAT}, and $z$ is the redshift of the observed TDE. The limiting luminosity distance for a specific TDE depends on the redshift $z$, the TDE's age $t$, the SMBH mass $M_\bullet$, and the stellar mass $M_\star$ involved. Additionally, for volume calculations we use the comoving distance:
\begin{equation}
    D_{\rm C}=\frac{D_{\rm L}}{1+z}=\frac{c}{H_0}\int_0^z{\frac{dz'}{\sqrt{\Omega_{\rm m}(1+z')^3+\Omega_\Lambda}}}
    \label{eq: comoving distance}
\end{equation}
where $H_0=69.6\, \rm{km\,s^{-1}\,Mpc^{-1}}$ is the Hubble constant, and $\Omega_{\rm m}=0.286$ and $\Omega_\Lambda=0.714$ are the matter density and dark energy density, respectively \citep{Bennett+2014}. 

A fossil TDE will eventually become undetectable, either when it becomes optically thin or when $R_{\rm UV}$ shrinks to the ISCO, whichever occurs first. Additionally, the TDE luminosity will eventually decline too slowly to be identifiable as a transient. We considered two factors that limit the detectability of luminosity evolution: the statistical uncertainty in magnitude measurements and the photometric stability of \textit{ULTRASAT}. 

The statistical uncertainty includes shot noise from both the source and the background, as well as instrumental noise. A conservative estimate of these noise sources is provided in the \textit{ULTRASAT} scientific paper \citep{Shvartzvald+2024}. In our calculations, we assumed a normal distribution for the point spread function (PSF) when the number of pixels in the seeing disk is $N_{\rm{pix}}=4\pi\left(\text{FWHM}/2.35/\text{pixScale}\right)^2$, where FWHM$=8.3''$ is the PSF full width at half maximum, and pixScale$=5.4''/$pix. The PSF photometric efficiency is assumed to be $86\%$, and the throughput is $25\%$. To estimate the effective aperture area of the telescope, we use the fact that a limiting magnitude of $m_{\rm{AB}}=22.5$ corresponds to a $5\sigma$ detection in a $900$~s exposure. To improve the signal-to-noise ratio (SNR), we assume the co-adding of multiple images taken every four days over a six-month period. We consider a fossil TDE as identifiable if there is a $3\sigma$ detection of time evolution between two six-month co-added exposures separated by a year. Additionally, we consider $1\sigma$ and $2\sigma$ thresholds for constructing a lower quality fossil candidate library, which could be verified with followup investigations by other instruments. 

Another constraint on the detectability of evolution is the photometric stability of \textit{ULTRASAT}. This limit  will likely allow us to detect luminosity changes of $\Delta L=1\%$ over a year \citep{Shvartzvald+2024}, though we also consider a more conservative estimate\footnote{This likely represents a worst-case scenario for {\it ULTRASAT}.  Eran Ofek, private communication.} of $\Delta L=5\%$. 
 
\begin{figure}
    \includegraphics[width=85mm]{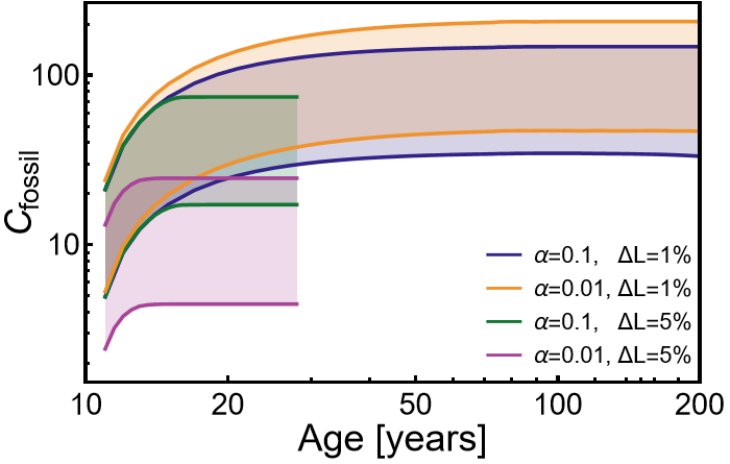}
    \par 
    \includegraphics[width=85mm]{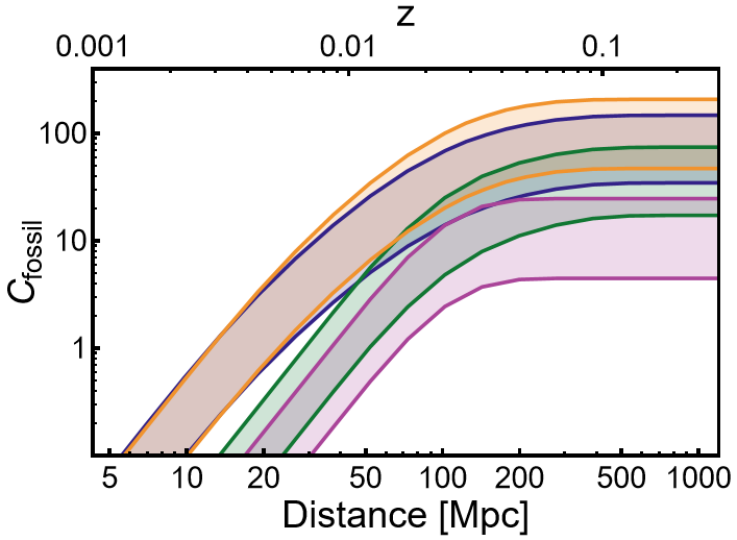}\\\par
    \includegraphics[width=85mm]{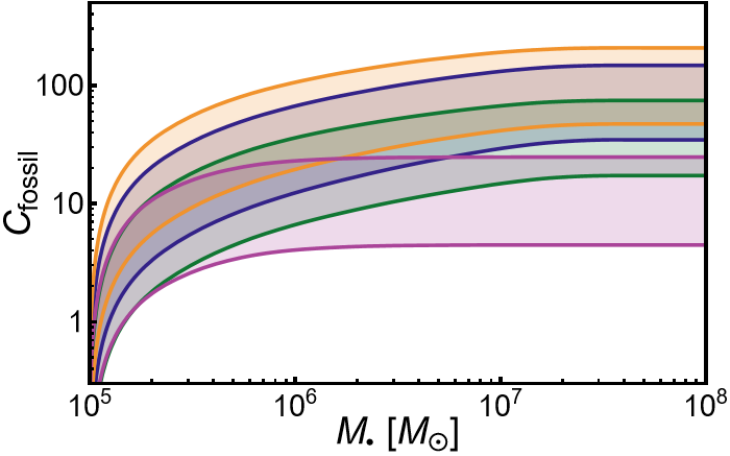}\par
\caption{Cumulative distribution functions showing the number of {\it ULTRASAT}-observable fossil TDEs as a function of their age (top) \red{and}\blue{,} their comoving distance and redshift (\red{bottom}\blue{middle}) \blue{and SMBH mass (bottom)}. The limiting NUV AB magnitude in these estimations is $m_{\rm AB}=22.5$. The results are shown for a $3\sigma$ detection threshold for the minimum detectable luminosity change over 1 year, and with two different photometric stability limits: $\Delta L=1\%$ (blue and orange), and $\Delta L=5\%$ (green and pink). The predicted number of {\it ULTRASAT} fossils are presented for two different estimates of disk effective viscosity: $\alpha=0.01$ (orange and pink) and $\alpha=0.1$ (blue and green). In these calculations, we considered both the empirical TDE rate from the ZTF sample of TDEs (lower curves; \citealt{Yao+2023}) and the dynamically predicted rate from loss cone theory (higher curves; \citealt{Stone+2016}).  Under our most pessimistic assumptions (low event rate, low $\alpha$, poor photometric stability), {\it ULTRASAT} will find $\sim$ few TDE fossils; under our most optimistic assumptions (high event rate, high $\alpha$, good photometric stability), \red{$\sim 340$}\blue{$\sim 210$}.}
\label{fig: fossils vs age}
\end{figure}

The number of detectable fossil TDEs can be estimated using:
\begin{equation}
\begin{aligned}
    N_{\rm fossils}&\approx\int_{0.08M_\odot}^{1M_\odot}dM_\star\int_{10^5M_\odot}^{M_{\rm Hills}}dM_\bullet\int_{10\,\rm years}^{200\,\rm years}dt\int_0^{z_{\rm max}}dz\\
    &\red{\dot{n}_{\rm TDE}}\blue{\dot{N}_{\rm TDE}}(M_\bullet)\frac{dV(z,t)}{dz}\frac{d\red{N_\bullet}\blue{n_\bullet}(z,M_\bullet)}{dM_\bullet}\frac{dN_\star(M_\star)}{dM_\star}\\
    &\times\frac{\Delta\Omega}{4\pi}H(z,t,M_\bullet,M_\star)
\end{aligned}
\end{equation}
where we define the minimum age for a TDE to be classified as a fossil as $10$ years. \blue{Even though TDEs remain bright for many centuries, identifying them as fossil TDEs is more challenging because their classification also requires observable time evolution. Therefore, we limit our integration to $200$ years, which, as shown in our results in \cref{fig: fossils vs age}, is a reasonable choice.} \red{$\dot{n}_{\rm TDE}$} \blue{$\dot{N}_{\rm TDE}$} is the \red{volumetric}\blue{per-galaxy} TDE rate, which can be estimated from Zwicky Transient Facility (ZTF) observations \blue{($\dot{N}_{\rm TDE}(M_\bullet)=4.1\times10^{-5}\left(\frac{M_\bullet}{10^6M_\odot}\right)^{-0.25}$ yr$^{-1}$ gal$^{-1}$}; \citealt{Yao+2023}) or theoretical estimations \blue{($\dot{N}_{\rm TDE}(M_\bullet)=1.9\times10^{-4}\left(\frac{M_\bullet}{10^6M_\odot}\right)^{-0.404}$ yr$^{-1}$ gal$^{-1}$}; \citealt{Stone+2016}). $V=\frac{4\pi}{3}D_{\rm C}^3$ is the comoving volume, $\frac{d\red{N_\bullet}\blue{n_\bullet}}{dM_\bullet}$ is the \blue{volumetric} SMBH mass function \citep{Shankar+2009}, and $\frac{dN_\star}{dM_\star}$ is the Kroupa present-day mass function for an old nuclear population:
\begin{equation}
    \frac{dN_\star}{dM_\star}=
    \begin{cases}
        \frac{0.28}{M_\odot}\left(\frac{M_\star}{M_\odot}\right)^{-1.3} & , 0.08 \leq \frac{M_\star}{M_\odot} \leq 0.5\\
        \frac{0.14}{M_\odot}\left(\frac{M_\star}{M_\odot}\right)^{-2.3} & , 0.5 \leq \frac{M_\star}{M_\odot} \leq 1\\
        0 & ,\text{otherwise}.
    \end{cases}
\end{equation}
To account for all the limiting factors described in the previous paragraph, we define a step function $H(z,t,M_\bullet,M_\star)$ that returns $1$ when the limiting conditions are satisfied and $0$ otherwise. 

The maximum SMBH mass, $M_{\rm Hills}(M_\star)$, corresponds to the Hills mass \citep{Hills1975}, determined by the requirement that the tidal radius, $r_t$, must be outside the innermost bound circular orbit (IBCO). For simplicity, we use the Schwarzschild IBCO in our calculations. \blue{We use $10^5 M_\odot$ as the lower integration limit for the SMBH mass, as this is the smallest value in the SMBH mass function from \cite{Shankar+2009}. Additionally, detectable fossil TDEs are limited by their luminosity, which increases with SMBH mass, as shown in \cref{fig: light curves- RUV}.}

The maximum redshift, $z_{\rm max}(t,M_\bullet,M_\star)$, at which a fossil TDE can be detected  is calculated using \cref{eq: luminosity distance,eq: comoving distance}. 
Furthermore, \textit{ULTRASAT} will not survey the entire sky but only a fraction of it. In its low-cadence mode, \textit{ULTRASAT}'s sky coverage is approximately $\Delta\Omega\approx 6800\, \rm{deg}^2$ \citep{Shvartzvald+2024}.  

The results of our estimate for the number of observed fossil TDEs with $3\sigma$ detection are presented in \cref{fig: fossils vs age}, which shows the cumulative number of fossil TDEs as a function of their age and their distance. The results are presented for different \textit{ULTRASAT} photometric stability thresholds, Shakura-Sunyaev $\alpha$ prefactor values, and TDE rates. 

Additionally, by assuming less stringent detection limits, we can imagine building a low-specificity library of fossil candidates that could be verified through future observations  with other instruments. Therefore, in \cref{table: fossils}, we present the total number of fossil TDEs and candidates that \textit{ULTRASAT} could detect under different detection thresholds for luminosity evolution. 

\begin{table}[h]
    \centering
    \begin{tabular}{|c||c|c|c|c|}
     \multicolumn{5}{c}{(a) $\alpha=0.1$}\\
     \hline 
     & \multicolumn{2}{c}{$\Delta L=1\%$} & \multicolumn{2}{|c|}{$\Delta L=5\%$} \\
     \hline 
     & Observations & Theory & Observations & Theory\\
     \hline 
     $3\sigma$& \red{66}\blue{34}  & \red{281}\blue{147} &  \red{29}\blue{17}  & \red{130}\blue{74}\\
     \hline
     $2\sigma$& \red{120}\blue{58} & \red{515}\blue{250} &  \red{48}\blue{25}  & \red{215}\blue{108} \\
     \hline
     $1\sigma$& \red{212}\blue{93} & \red{911}\blue{396} &  \red{51}\blue{26}  & \red{227}\blue{111} \\
     \hline
    \end{tabular}
    \bigskip
    
    \begin{tabular}{|c||c|c|c|c|}
     \multicolumn{5}{c}{(b) $\alpha=0.01$}\\
     \hline 
     & \multicolumn{2}{c}{$\Delta L=1\%$} & \multicolumn{2}{|c|}{$\Delta L=5\%$} \\
     \hline 
     & Observations & Theory & Observations & Theory\\
     \hline 
     $3\sigma$& \red{76}\blue{47} &   \red{338}\blue{207}  &  \red{4}\blue{4}   &   \red{22}\blue{25}  \\
     \hline
     $2\sigma$& \red{151}\blue{92} &  \red{669}\blue{401} &   \red{6}\blue{7}   &   \red{35}\blue{40}  \\
     \hline
     $1\sigma$& \red{345}\blue{196} & \red{1484}\blue{833} &   \red{7}\blue{8}   &   \red{38}\blue{43}  \\
     \hline
    \end{tabular}
    \caption{Total number of fossil TDEs and fossil candidates observable by {\it ULTRASAT}. We use the same parameters as in \cref{fig: fossils vs age}, but with different detection thresholds for the luminosity evolution ($3\sigma$, $2\sigma$, $1\sigma$).  ``Observations'' and ``Theory'' refer to whether we use (low) observationally-motivated TDE rates or (high) theoretically-motivated TDE rates.  In Table (a) we consider rapidly evolving disks ($\alpha=0.1$) and in Table (b) we consider slowly evolving ones ($\alpha=0.01$).
    }
    \label{table: fossils}
\end{table}

\cref{fig: fossils vs age} and \cref{table: fossils} show that for the $3\sigma$ detection threshold with the conservative threshold of $\Delta L=5\%$, the most pessimistic combination of assumptions ($\alpha=0.01$ and a TDE rate from ZTF) predicts the detection of a few fossil TDEs at $\sim 10$ years old, while the most optimistic scenario (with $\alpha=0.1$ and a TDE rate from theory) predicts the detection of \red{$\sim 130$}\blue{75} fossil TDEs. For the less conservative threshold of $\Delta L=1\%$, \textit{ULTRASAT} is expected to detect \red{$\sim 70$}\blue{35} fossil TDEs from decades ago in the most pessimistic case (with $\alpha=0.1$ and a TDE rate from ZTF), and up to \red{$\sim 340$}\blue{210} in the most optimistic case (with $\alpha=0.01$ and a TDE rate from theory). 

There are qualitative differences between the low photometric stability ($\Delta L=5\%$) case and the high photometric stability ($\Delta L=1\%$) case.  When $\Delta L$ is large, more fossils are detected as $\alpha$ increases, because the limiting factor for detection is observable time evolution.  When $\Delta L$ is small, more fossils are detected as $\alpha$ decreases, because here the limiting factor for detection is UV flux.  In the $\Delta L=5\%$ case, statistical uncertainties are generally unimportant, but they are dominant for $\Delta L = 1\%$. \blue{Additionally, in all cases, we find that most detectable fossil TDEs originate from low-mass SMBHs, indicating that the limiting factor for detectability is the evolution rate rather than the source luminosity.}

For less stringent detection thresholds ($2\sigma$ and $1\sigma$), we can find larger number of fossil candidates. This means that \textit{ULTRASAT} offers a promising opportunity to detect both fossil TDEs and fossil candidates even decades after the peak of the light curve.

\subsection{TDE-QPEs}
Almost all galaxies above a certain size host a SMBH at their center, surrounded by a dense cluster of stars and compact objects. This dense nucleus is an excellent environment for various dynamical processes, such as TDEs, as well as the formation of extreme-mass ratio inspirals (EMRIs). An EMRI occurs when a stellar-mass compact object spirals toward the central SMBH through the emission of gravitational waves (GWs) \citep{Amaro-Seoane2018}. The compact object can approach very close to the SMBH while orbital energy is radiated away causing its semi-major axis to shrink. Recent studies suggest that the combination of a TDE with an EMRI may be the origin of a newly observed class of X-ray transients known as ``quasi-periodic eruptions'' (QPEs) \citep{Linial&Metzger2023, Franchini+2023}.    

QPEs are strong and short X-ray bursts that recur every few hours and originate near central SMBHs \citep{Miniutti+2019,Giustini+2020,Arcodia+2021,Chakraborty+2021,Arcodia+2024,Guolo+2024,Nicholl+2024}. It is currently unknown what triggers these events and how long they last, but any viable explanation for the QPE phenomenon requires a mechanism capable of abruptly and quasi-periodically feeding the innermost region of a relatively low-mass SMBH.

One class of mechanisms for the QPEs are the collisions between an orbiting secondary object and an accretion disk around a primary SMBH \citep{Sukova+2021,Xian+2021,Linial&Metzger2023,Tagawa&Haiman2023,Franchini+2023,Linial&Metzger2024}. When an EMRI passes twice per orbit through the accretion disk it perturbs the disk and the disturbances can potentially trigger the observed QPEs. The QPE luminosity can be produced by shocked disk material ejection, bow shocks in the disk, or by enhanced accretion rate. Some of the most promising models suggest that the source of the accretion disk is from a TDE \cite{Linial&Metzger2023, Franchini+2023} and the interactions begin when the spreading disk from the TDE has sufficiently radially extended until it meets an EMRI.  

This particular idea is supported by recent observations of a known TDE\footnote{But see also \citet{Chakraborty+2021,Guolo+2025,Wevers+2025}.} which has started to exhibit X-ray QPEs \citep{Nicholl+2024}. The TDE AT2019qiz was first discovered by ZTF, and over several months, its UV and optical luminosity declined until reaching its plateau phase. Approximately $1500$ days after its first optical detection, the \textit{Chandra} X-ray Observatory found repeating sharp increases in the X-ray luminosity. The mean recurrence time between peaks is $48.4\pm 0.3$ hours, with typical peak durations of $8-10$ hours, consistent with previous QPE observations. 

This observation is the first TDE to exhibit clear QPEs later in its lifetime, suggesting that the QPEs originate from a spreading disk formed by the TDE. Therefore, we want to check how long it takes the disk to spread until it encounters a secondary object.  

For simplicity, we assume a BH with mass $m_\bullet$ in a \blue{Keplerian} circular orbit as the secondary object in our calculations. \blue{If the secondary object crosses the accretion disk twice per orbit, the QPE recurrence time is given by $t_{\rm rec}=\pi\sqrt{\frac{r_{\rm sit}^3}{GM_\bullet}}$ where $r_{\rm sit}$ is the radius at which the secondary object is likely to be located. This radius corresponds to the radius that a TDE needs to spread in order for QPEs to begin. By measuring $t_{\rm rec}$ we can calculate $r_{\rm sit}$ and then calculate $t_{\rm on}$, the time it takes for the TDE to spread to the radius $r_{\rm sit}$ and the QPEs to turn on. } \red{The EMRI rate is highly uncertain \citep{Babak2017}, so we explore a range of possible values: $\dot{N}_{\rm EMRI}=10^{-6}-10^{-8}\, \rm{yr}^{-1}\rm{gal}^{-1}$ \citep{Gair+2004}. Since the BH's semi-major axis contracts due to the emission of GW, the EMRI rate is approximately $\dot{N}_{\rm EMRI}\approx 1/t_{\rm GW}$, where $t_{\rm GW}$ is the GW inspiral time: the time it takes the BH to evolve from an initial semi-major axis to merger. For a circular orbit with an initial semi-major axis $a_0$ the GW timescale is given by:  
\begin{equation}
   t_{\rm GW}= \frac{5c^5a_0^4}{256G^3M_\bullet m_\bullet(M_\bullet+m_\bullet)}.
\end{equation}
So, by knowing the EMRI rate, we can calculate the radius $r_{\rm sit}\equiv a_0$ where an EMRI is likely to be found. This radius corresponds to the radius that a TDE needs to spread in order to start exhibiting QPEs.  }

\begin{figure}
\includegraphics[width=85mm]{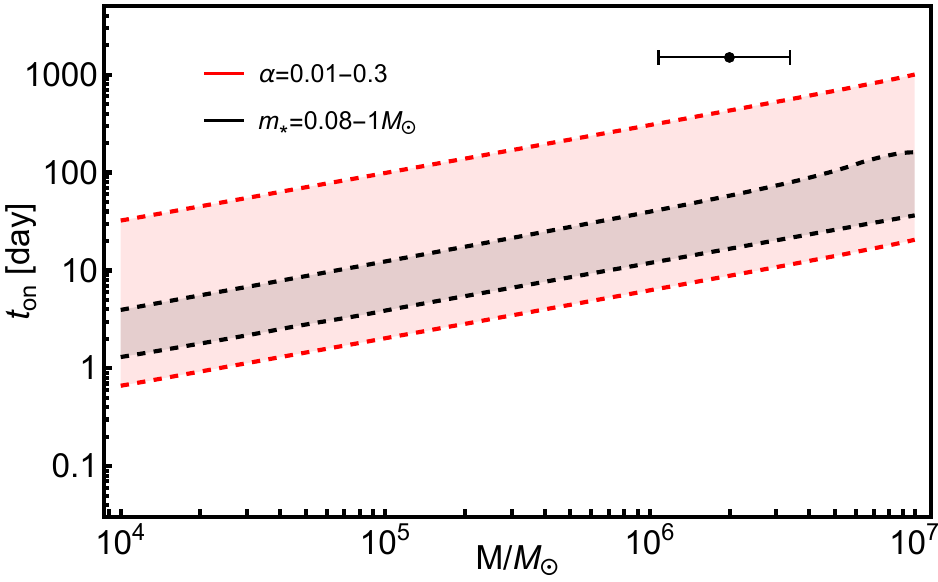}
\includegraphics[width=85mm]{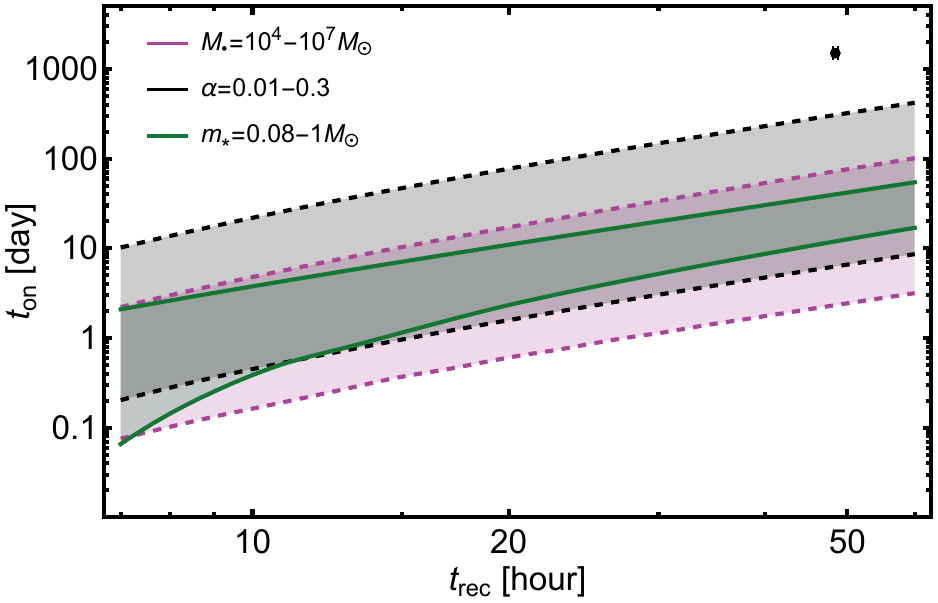}
\caption{The time it takes for the QPEs to turn on after the TDE burst as a function of the SMBH mass (\textit{Top}) and recurrence time (\textit{Bottom}). The secondary object is a BH in a circular orbit with mass $m_\bullet$. The default parameters are $m_\bullet=10M_\odot$, \red{$\dot{N}_{\rm EMRI}=10^{-7}\, \text{yr}^{-1}\text{gal}^{-1}$}\blue{$t_{\rm rec}=48.4$ hours}, $\alpha=0.1$, $M_\bullet=10^6 M_\odot$, and $M_\star=0.3M_\odot$. \textit{Chandra} observed the beginning of the QPEs $\sim 1500$ days after AT2019qiz was first detected, around a SMBH with $\log_{10} M_\bullet/M_\odot=6.3^{+0.3}_{-0.2}$, with a mean recurrence time between peaks of $48.4\pm 0.3$ hours \citep{Nicholl+2024}. The observed time to exhibit QPEs of AT2019qiz is longer than the expected time from our model.}
\label{fig: QPE-TDE}
\end{figure}

In \cref{fig: QPE-TDE}, we show the time $t_{\rm on}$ \red{it takes for the TDE to spread to the radius $r_{\rm sit}$ and the QPEs to turn on,} as a function of the SMBH mass and the QPEs mean recurrence time. The results are shown for various parameter ranges of the \red{EMRI rate,} secondary mass, the alpha parameter, and the mass of the disrupted star. In all cases, the QPEs turn on before the time when \textit{Chandra} observed the first QPE in AT2019qiz. \cite{Nicholl+2024} were unable to constrain exactly when the QPEs began in AT2019qiz. Therefore, our results suggest that the QPEs may have started before the first observed X-ray burst, or alternatively could originate from a different mechanism. However, it is plausible that not all disk-orbiter collisions produce detectable X-ray flares, and additional conditions beyond the physical extent of the disk are required. The ``TDE+EMRI=QPE'' model could, in principle, be tested with more continuous X-ray monitoring of TDEs over a range timescales as shown in Fig. \ref{fig: QPE-TDE}.

\section{Conclusions} \label{sec: conclusions}

Standard accretion disk models are viscously and thermally unstable in the inner regions of the disk, where radiation pressure dominates over gas pressure \citep{Pringle+1973,Shakura&Sunyaev1976,Lightman1974,Piran1978}. However, these instabilities are not so far observed in TDEs. A common theoretical approach to stabilize the disk is to modify the viscosity parametrization \citep{Sakimoto&Coroniti1981,Taam&Lin1984,Mummery2019}, though the most widely used parametrizations are often ad hoc.

In this paper, we examine the $\alpha$-disk model \citep{Shakura&Sunyaev1973}, under the assumption that magnetic fields are the dominant source of pressure. Given the presence of MRI, we expect the disk to be highly magnetized. Previous research has shown that magnetic fields can indeed stabilize the disk against radiation pressure instabilities \citep{Begelman&Pringle2007,Sadowski2016,Jiang+2019,Huang+2023,Kaur+2023}.

We therefore compared the commonly used linear viscosity parametrization with the magnetized $\alpha$-viscosity model. The linear viscosity is widely used because it stabilizes the disk and permits analytic solutions to the surface density evolution equation. However, for the magnetized $\alpha$-disk, the evolution equation becomes a nonlinear diffusion equation, which cannot be solved analytically. It can, however, be solved numerically, and there is an analytical self-similar solution that is valid at late times.  

We examined various values for the power law of the linear viscosity and found that, \red{although the surface densities look quite different,} \blue{both the surface density at late times and } the light curves may look very similar when the power law of the linear viscosity is chosen carefully. \blue{Specifically, the surface density at late times matches well when the linear viscosity follows a power law with $\mu_{\rm l}=5/9$. Additionally, we} \red{We} found that the time evolution of the UV luminosity for the magnetized viscosity \red{is} \blue{scales as} $L_{\rm m,UV}\propto t^{-5/6}$, and that the light curves for the linear viscosity fortuitously follow the same time evolution when the power law is set to $\mu_{\rm l}=0$.

We fit our model to the well-observed optical and UV light curves of the TDE ASASSN-14li. The evolution of its light curves indicates that the viscous spreading in ASASSN-14li is slow, leading previous studies to propose mechanisms such as angular momentum loss in a magnetized wind \citep{Wen+2023} to limit the disk’s outer spreading. Our results show that the best-fit Shakura-Sunyaev $\alpha$ parameter falls within a theoretically reasonable range, requiring no additional physical effects.

We also noted that the UV light curves of the magnetized disk evolve slowly and TDEs may remain bright for decades. We propose that it may be possible to detect fossil TDEs that erupted a long time ago using \textit{ULTRASAT} or other future wide-field UV sky surveys. To detect fossil TDEs, we need the source to be bright enough and to evolve quickly enough for a detectable change in brightness over the course of a year. Since the late stages of a TDE are much simpler than the early stages, fossil TDEs could provide opportunities to measure various SMBH parameters. We estimated the number of fossil TDEs that \textit{ULTRASAT} might be able to detect, and found that it might detect hundreds of fossil TDEs that erupted decades ago. 

In addition, recent observations have detected a TDE that began to exhibit QPEs several years after the first TDE observation. Although the exact mechanism behind QPEs remains uncertain, this observation suggests a link between QPEs and TDE accretion disks. One proposed mechanism involves a TDE disk expanding until it eventually encounters an EMRI, with the EMRI passing through the disk twice per orbit and producing X-ray flares. We estimated the time it takes for the magnetized disk to spread enough to encounter an EMRI and found that the QPEs should have started earlier than the observed QPEs.  

In this paper, we studied accretion disks in a 1D framework, assuming a single pressure component due to magnetic fields. In reality, radiation pressure may also be non-negligible \citep{Jiang+2019} and should be considered alongside magnetic pressure for a more comprehensive model. Furthermore, while we used an approximate estimation for the magnetic field strengths based on an MRI saturation criterion \citep{Pessah&Psaltis2005,Begelman&Pringle2007} which is consistent with recent radiation MHD simulations \citep{Jiang+2019,Mishra+2022,Huang+2023}, alternative saturation criteria \citep{Oda+2009,Begelman&Armitage2023} and other MHD instabilities \citep{Das+2018} deserve further investigation and could potentially influence our results.
Nevertheless, the magnetized viscosity parametrization  adopted in this work remains both simple and more physically motivated compared to other commonly used parametrizations.

In future work, we hope to develop a time-dependent model that combines gas, radiation, and magnetic pressures. By exploring various ratios between these components, we could study and better understand variability in different accretion disk systems, such as QPEs and TDEs. While magnetic fields appear to be a promising mechanism for stabilizing disks against radiation pressure instabilities, the exact details of this process remain uncertain and require further investigation through both analytical and numerical studies.

\begin{acknowledgments}
YA gratefully acknowledges the support of Azrieli fellowship.
YA and NCS gratefully acknowledge support from the Binational Science Foundation (grant Nos. 2019772 and 2020397) and the Israel Science Foundation (Individual Research Grant Nos. 2565/19 and 2414/23).  We thank Sjoert van Velzen for his assistance in providing late-time TDE light curves and Iair Arcavi, Eran Ofek, and Yossi Shvartzvald for discussions on the photometric stability and statistical uncertainties of {\it ULTRASAT}.  We also acknowledge helpful comments from Mitch Begelman, Agnieszka Janiuk, Itai Linial, Brian Metzger, Cole Miller, Peter Jonker, Sixiang Wen, and Jim Pringle on an earlier draft of the paper.

\end{acknowledgments}

\appendix
\section{Separability of Disk Equation}
\label{appendix: separability Emden-Fowler}
The evolution equation (Eq. \ref{eq:diffusion eq}) with a magnetized $\alpha$-viscosity (Eq. \ref{eq:viscosity}) is a nonlinear partial differential equation. Here we show that \cref{eq:diffusion eq} can be reduced to two ordinary differential equations: (i) a trivial time-dependent equation and (ii) a nonlinear Emden-Fowler equation.  To the best of our knowledge, this has not been demonstrated previously despite the wide applicability of Eq. \ref{eq:diffusion eq} in high energy astrophysics.  There are, however, more specific investigations of separability of Eq. \ref{eq:diffusion eq} in certain limits, such as self-similar solutions \citep{Rafikov2016}.

We consider a general double power law viscosity 
\begin{equation}
    \nu=\nu_0 R^n \Sigma^m 
\end{equation}
where for the magnetized $\alpha$-viscosity $m=2/7$ and $n=5/7$. We use the following transformation from \cite{Pringle1991}: $x\equiv R^{1/2}$, $\tau\equiv 3\nu_0t/4$ and $S(\tau,x)\equiv\Sigma R^{3/2}$. \cref{eq:diffusion eq} then becomes:
\begin{equation}
    \frac{\partial S}{\partial\tau}=\frac{\partial^2}{\partial x^2}(S^{m+1}x^{2n-3m-2})
    \label{eq-appendix: diffusion eq after transform}
\end{equation}
Assuming that $S$ is separable: 
\begin{equation}
    S(\tau,x)=T(\tau)X(x)
\end{equation}
\cref{eq-appendix: diffusion eq after transform} becomes
\begin{equation}
    \frac{T'}{T^{m+1}}=\frac{1}{X}\frac{d^2}{dx^2}\left[X^{m+1}x^{2n-3m-2}\right].
\end{equation}

The left-hand side depends only on $\tau$, and the right-hand side depends only on $x$. Therefore, each side must be equal the same constant $\lambda$:
\begin{equation}
\begin{cases}
    \frac{T'}{T^{m+1}}=\lambda\\
    \frac{1}{X}\frac{d^2}{dx^2}\left[X^{m+1}x^{2n-3m-2}\right]=\lambda.
\end{cases}
\end{equation}
$\lambda$ is a constant that must be negative to obtain physically meaningful solutions, ensuring the disk mass declines over time and enabling the formulation of two-point boundary conditions in the radial equation. 

The time-dependent equation is trivial and can be solved easily. For $m=0$, the solution to \cref{eq:diffusion eq} has been determined previously (e.g. \citealt{Metzger+2008}), so we focus here on the case $m\neq 0$. In this case, the solution is $T(\tau)=(-m\lambda\tau+C)^{-1/m}$ where $C$ is an arbitrary constant of integration. 

The spatial equation is a nonlinear ordinary differential equation, which through a change of variables $w(x)\equiv X^{m+1}x^{2n-3m-2}$ transforms into an Emden-Fowler\footnote{The Lane-Emden equation is a well-studied special case of the Emden-Fowler equation.} equation \citep{zaitsev&Polyanin2002}:
\begin{equation}
    \frac{d^2w}{dx^2}=\lambda w^{\frac{1}{m+1}}x^{\frac{3m-2n+2}{m+1}}. 
\end{equation}
For the magnetized $\alpha$-viscosity we employ, the power of $w$ is $7/9$, and the power of $x$ is $10/9$. Although exact solutions exist for specific power-law values \citep{zaitsev&Polyanin2002}, no exact solution has yet been found for these power-law indices. 

\begin{figure}
\centering
\includegraphics[width=100mm]{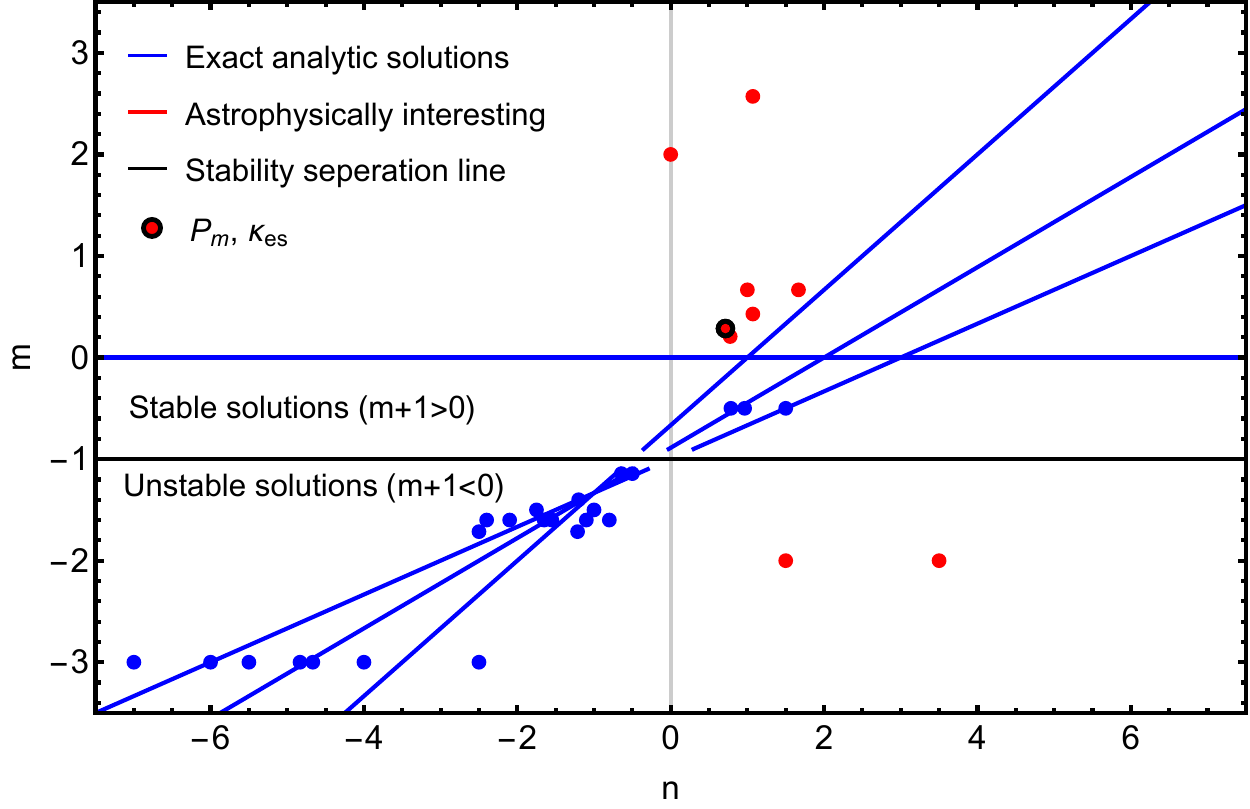}
\caption{A scatterplot of the viscosity power-law values. Exact solutions for the Emden-Fowler equation are shown as blue lines (continuous solutions) and blue dots (discrete solutions). Red dots represent astrophysically interesting values, corresponding to all permutations of  $P\in\left\{P_{\rm gas}, P_{\rm rad}, P_{\rm mag} \right\}$ and $\kappa\in\left\{\kappa_{\rm es}, \kappa_{\rm Kr}, \kappa_{\rm dust}\right\}$. The magnetized viscosity with $\kappa_{\rm es}$, which is the focus of this study, is highlighted as a bold red dot. All solutions above the black line are stable, while those below the line are unstable.  Currently, no exact solutions exist for the astrophysically interesting values.}
\label{fig: Emden-Fowler solutions}
\end{figure}

We have explored other astrophysically interesting power-law viscosity parametrizations. We calculated the values for every combination of gas, radiation, and magnetic pressures ($P\in\left\{P_{\rm gas}, P_{\rm rad}, P_{\rm mag} \right\}$), with every combination of electron scattering, Kramers, and dust opacities ($\kappa\in\left\{\kappa_{\rm es}, \kappa_{\rm Kr}, \kappa_{\rm dust}\right\}$). 
The power-law values of the exact solutions (both continuous and discrete), as well as those of astrophysical interest, are presented  in \cref{fig: Emden-Fowler solutions}. Unfortunately, no exact solutions currently exist for the values relevant to our study.  If in the future such solutions are found, they could provide greater physical intuition for the time evolution of viscously spreading accretion disks.

\section{Fitting the Self-Similar Solution}
\label{appendix: self-similar fit}
Using the self-similar solution is simpler and more efficient than numerically integrating the disk equations. In \S \ref{sec: disk model}, we demonstrated that the self-similar solution closely approximates the numerical solution at late times and large radii. In this appendix, we provide  a fitting function for the free parameter $\nu_{0,\rm{S}}$ in \cref{eq:similaritysol}. For each value of the $\alpha$ parameter, SMBH mass $M_\bullet$, and stellar mass $M_\star$ satisfying $M_\bullet<M_{\rm Hills}$, the self-similar free parameter can be determined using:
\begin{equation}
    \nu_{0,\rm S}=1.91\cdot 10^{15}\alpha^{1.08}\left(\frac{M_\star}{M_\odot}\right)^{0.40}\left(\frac{M_\bullet}{M_\odot}\right)^{0.19}\left(1+\frac{M_\bullet}{M_{\rm Hills}}\right)^{-0.46}.
\end{equation}

\begin{figure}
\centering
\includegraphics[width=85mm]{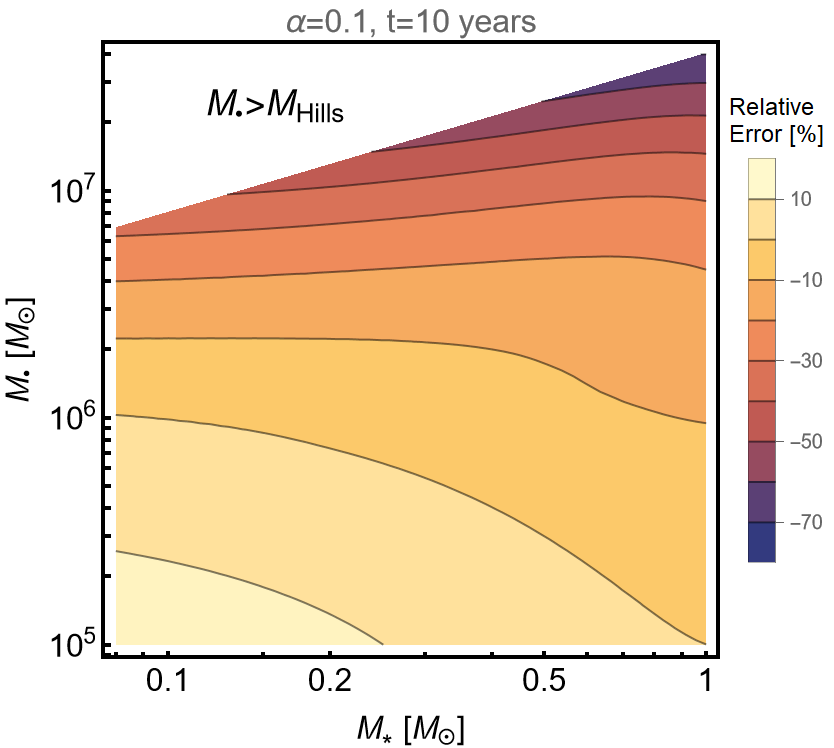}
\includegraphics[width=85mm]{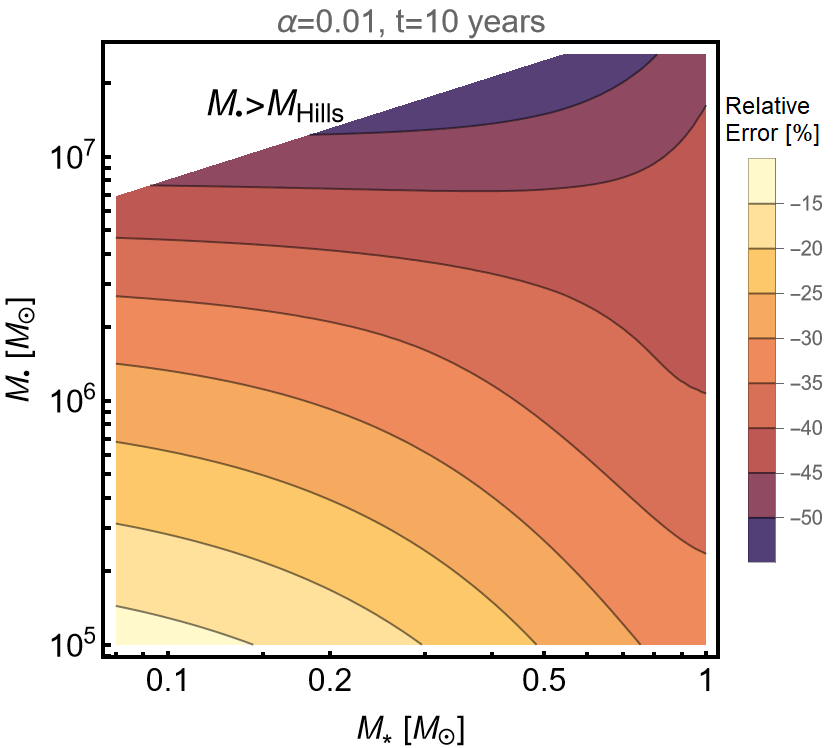}\par
\includegraphics[width=85mm]{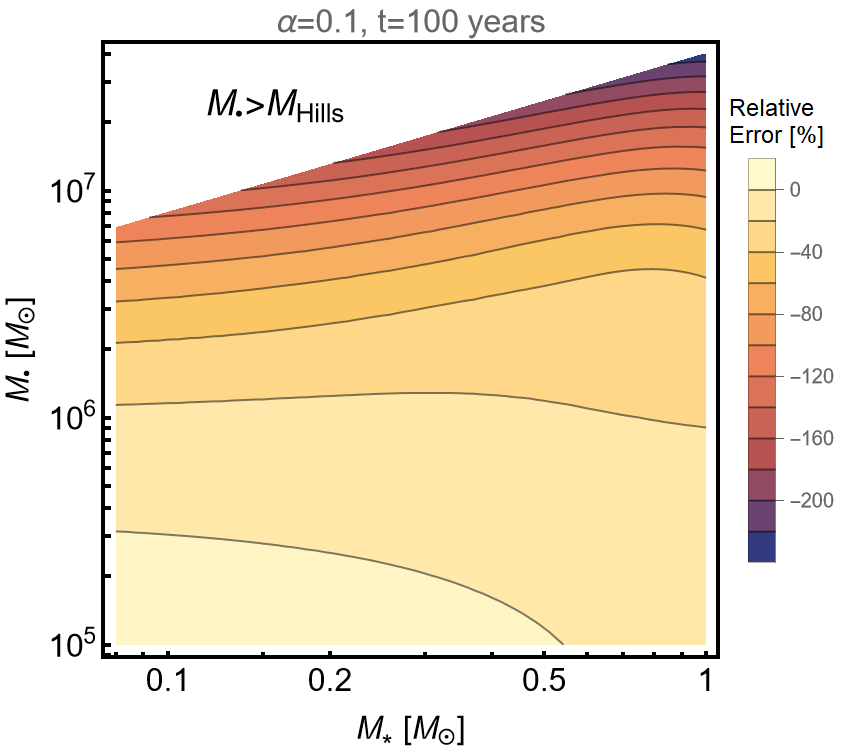}
\includegraphics[width=85mm]{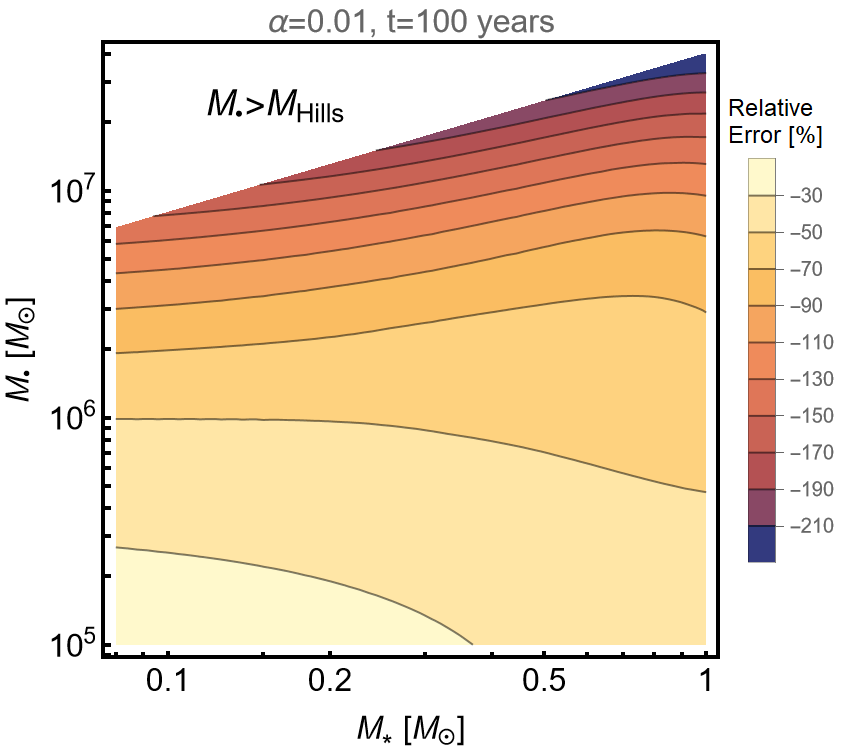}
\caption{The relative error between the UV light curves from the numerical solution  and the self-similar solution is shown as a function of the SMBH mass and stellar mass. The error is plotted for different $\alpha$ prefactor values ($\alpha=0.1$ and $\alpha=0.01$) and for different ages of the TDE ($t=10$ years and $t=100$ years). The white region corresponds to SMBH masses exceeding the Hills mass. The relative error is larger for masses near the Hills mass, while for other regions, the error remains below an order of magnitude. }
\label{fig: appendix self-similar error}
\end{figure}

\cref{fig: appendix self-similar error} shows the relative error in the  light curves at a UV wavelength of $\lambda=250 \rm{nm}$, between the numerical and the self-similar solutions: 
\begin{equation}
    \rm{Relative Error(t,\alpha,M_\bullet,M_\star)}=100\frac{L_{\red{\nu}\blue{f},\rm{num}}-L_{\red{\nu}\blue{f},\rm{S}}}{L_{\red{\nu}\blue{f},\rm{num}}}\,[\%]
\end{equation}
where $t$ is the age of the TDE, $L_{\red{\nu}\blue{f}, \rm{num}}$ is the spectral luminosity calculated from the numerical solution, and $L_{\red{\nu}\blue{f}, \rm{S}}$ is the spectral luminosity from the self-similar solution. The relative error is larger for SMBH masses near the Hills mass but remains below an order of magnitude in other regions.

\section{Comparison To Linear Viscosity with Additional Models}
\label{appendix: more linear viscosity comparison}

In this appendix, we present additional results.  We compare the magnetized $\alpha$-disk with the linear viscosity model, as was done in \S\ref{sec: linear viscosity}, for two additional scenarios. In the first scenario, we retain the Gaussian initial condition but normalize the linear viscosity prefactor, $\nu_{0,\rm l}$, in the initial conditions, rather than matching it to the magnetized disk after 1 year. Specifically, we set: $\nu_{0,\rm l}=\nu_{\rm mag}\left(r_{\rm c},\Sigma_0\right)/r_{\rm c}^{\mu_{\rm l}}$ where $\Sigma_0=\Sigma\left(R=r_{\rm c},t=0\right)$ is the initial surface density at the peak of the Gaussian. We compute results for various  power-law values $\mu_{\rm l}=\left\{1\blue{,5/9},0,-3/2\right\}$.

In the second scenario, we add a source function to  \cref{eq:diffusion eq} instead of using a Gaussian initial condition (while retaining the original normalization $\nu_{0,\rm{l}}$). This approach reflects the slow accumulation of mass from the bound debris into the accretion disk, at a rate given by Eq. \ref{eq:fallback}. For this case, we compute results for $\mu_{\rm l}=\left\{3/2\blue{,5/9},0,-3/2\right\}$. 

\begin{figure}
\centering
\includegraphics[width=85mm]{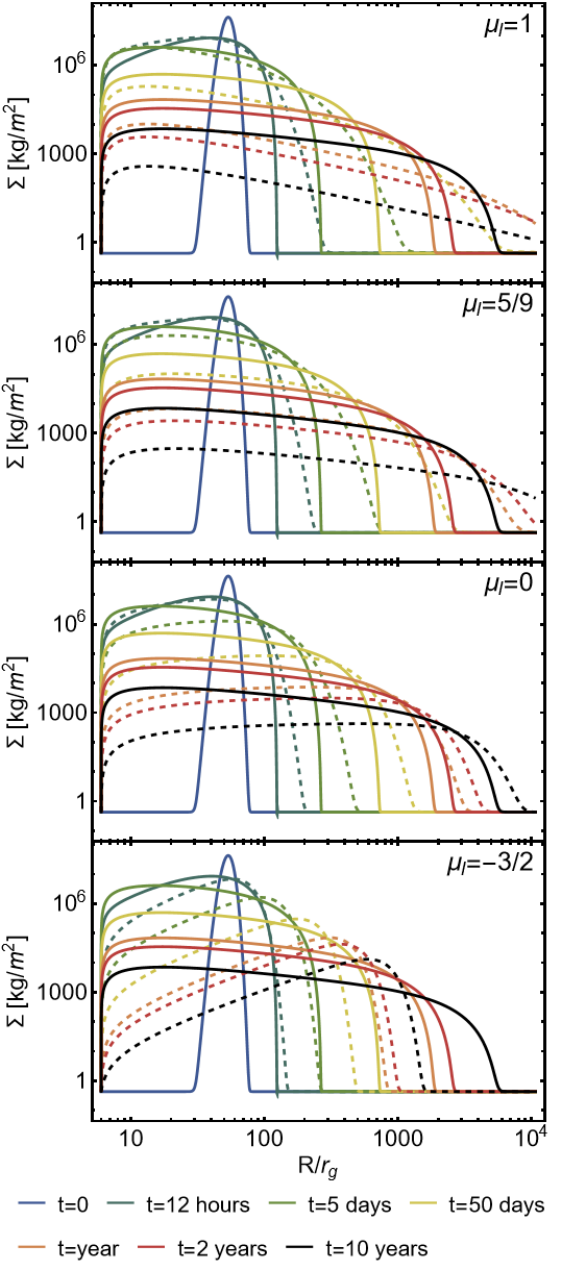}
\includegraphics[width=85mm]{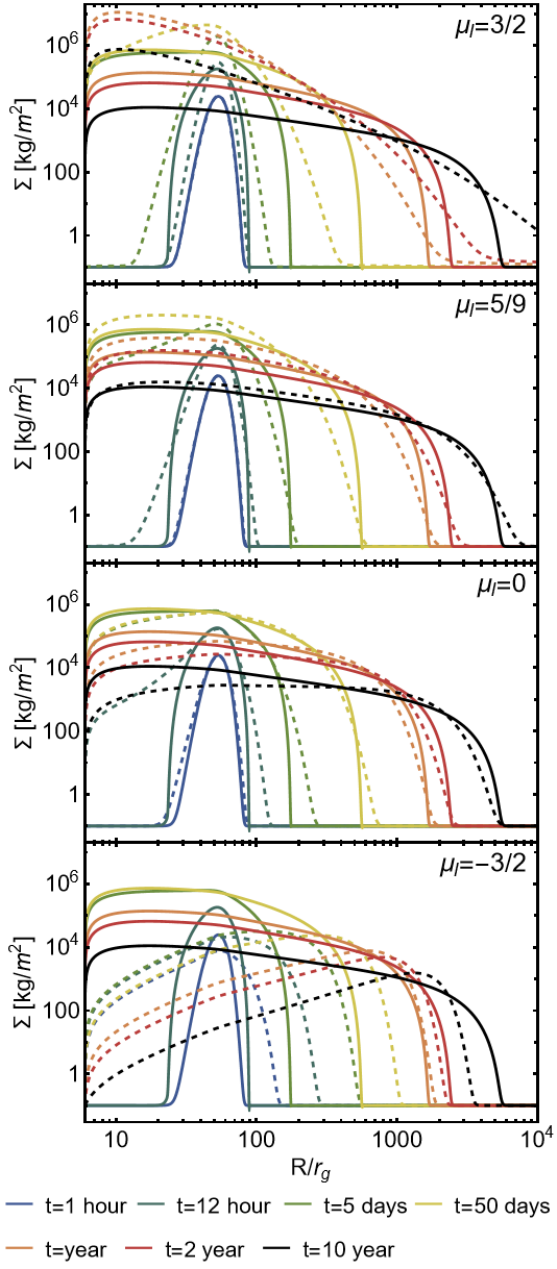}
\caption{Surface densities plotted against radius at different times
for the magnetized $\alpha$-viscosity (solid) and the linear viscosity (dashed), similar to \cref{fig:linear-mag comparison surface density}. The first case (top) assumes a Gaussian initial condition, with the normalization factor of the linear viscosity parametrization, $\nu_{0,\rm {l}}$, chosen to match the magnetized viscosity at the start time. The second case (bottom) includes a source term in the diffusion equation (Eq. \ref{eq:diffusion eq}) instead of a Gaussian initial condition. }
\label{fig: appendix Sigma vs R}
\end{figure}

\begin{figure}
\centering
\includegraphics[width=58mm]{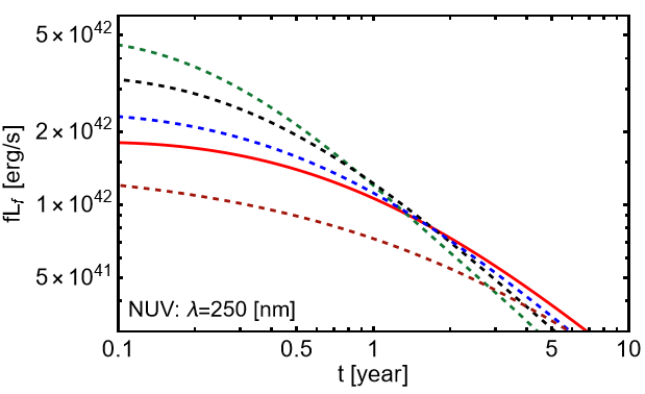}
\includegraphics[width=58mm]{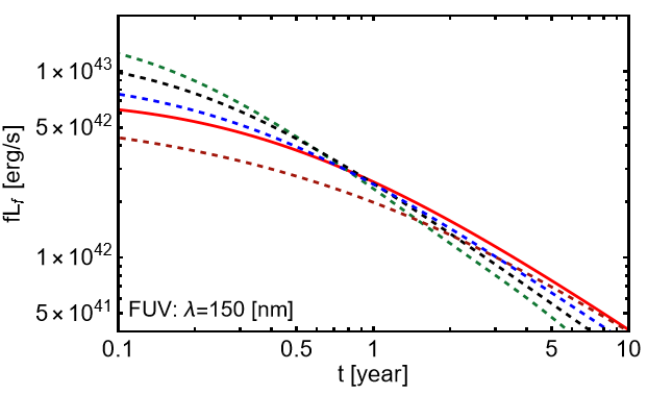}
\includegraphics[width=58mm]{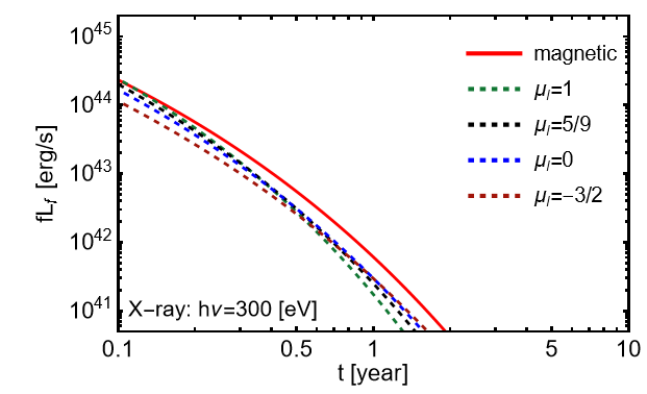}\par
\includegraphics[width=58mm]{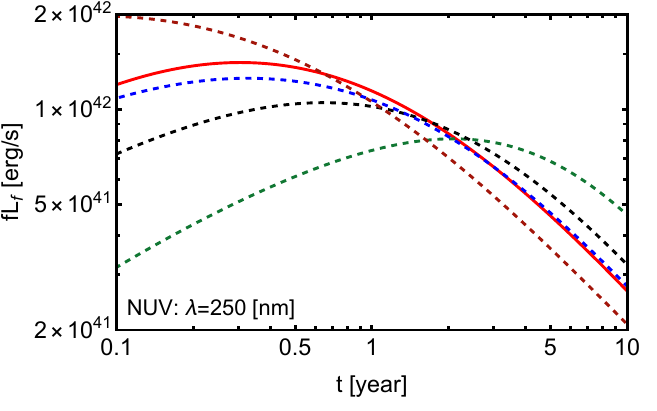}
\includegraphics[width=58mm]{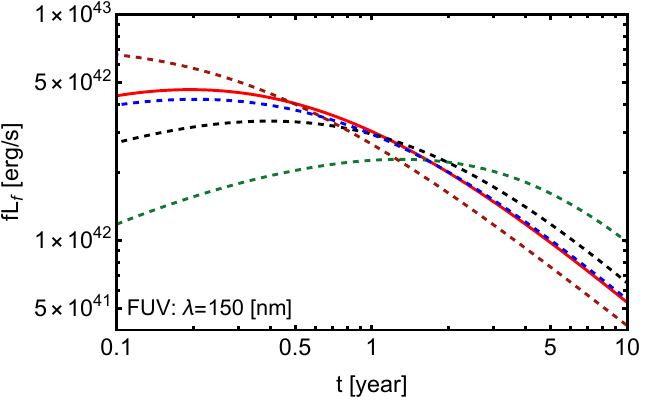}
\includegraphics[width=58mm]{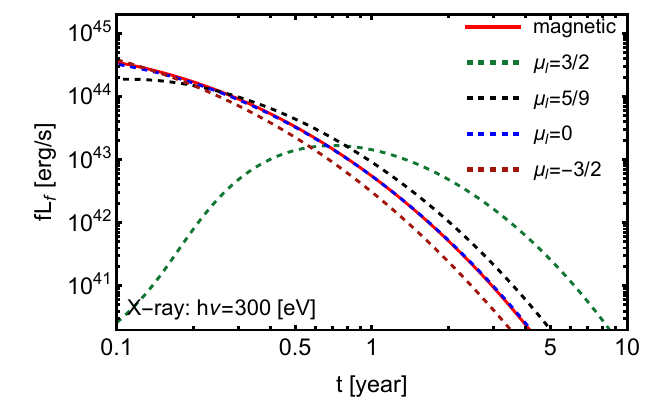}
\caption{Light curves for near-UV, far-UV and X-ray bands
(left to right) with the magnetized $\alpha$-viscosity (red) and the linear viscosity (dashed black, blue, brown), similar to \cref{fig:linear-mag comparison light curve}. The first case (top) assumes a Gaussian initial condition, with the normalization factor of the linear viscosity parametrization, $\nu_{0,\rm {l}}$, chosen to match the magnetized viscosity at the initial condition. The second case (bottom) includes a source term in the diffusion equation (Eq. \ref{eq:diffusion eq}) instead of a Gaussian initial condition.}
\label{fig: appendix light curves}
\end{figure}

\begin{figure}
\centering
\includegraphics[width=85mm]{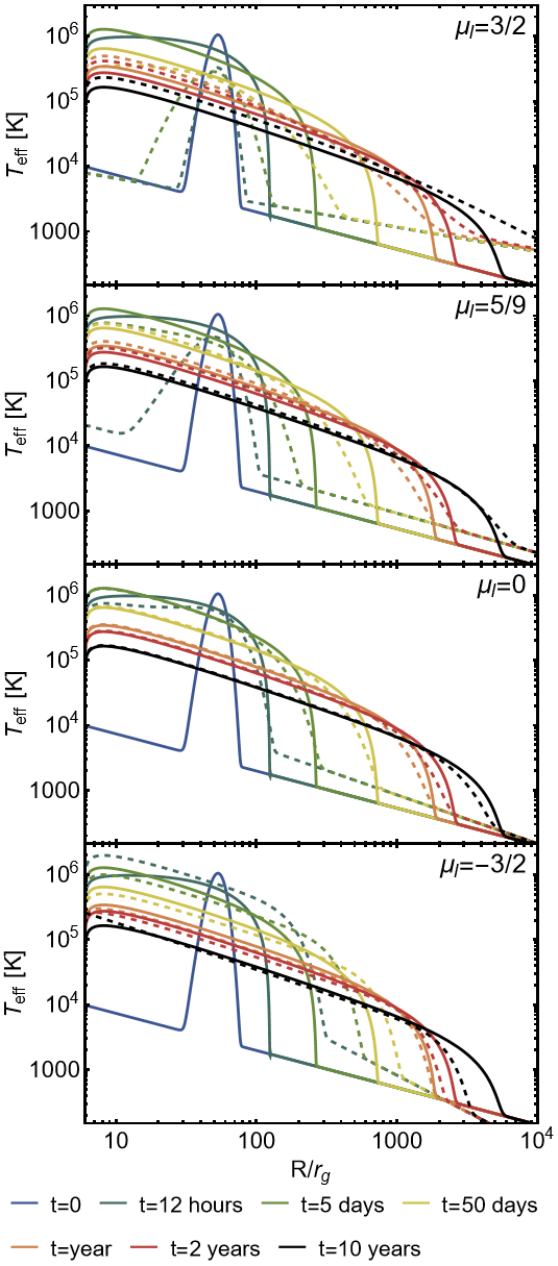}
\includegraphics[width=85mm]{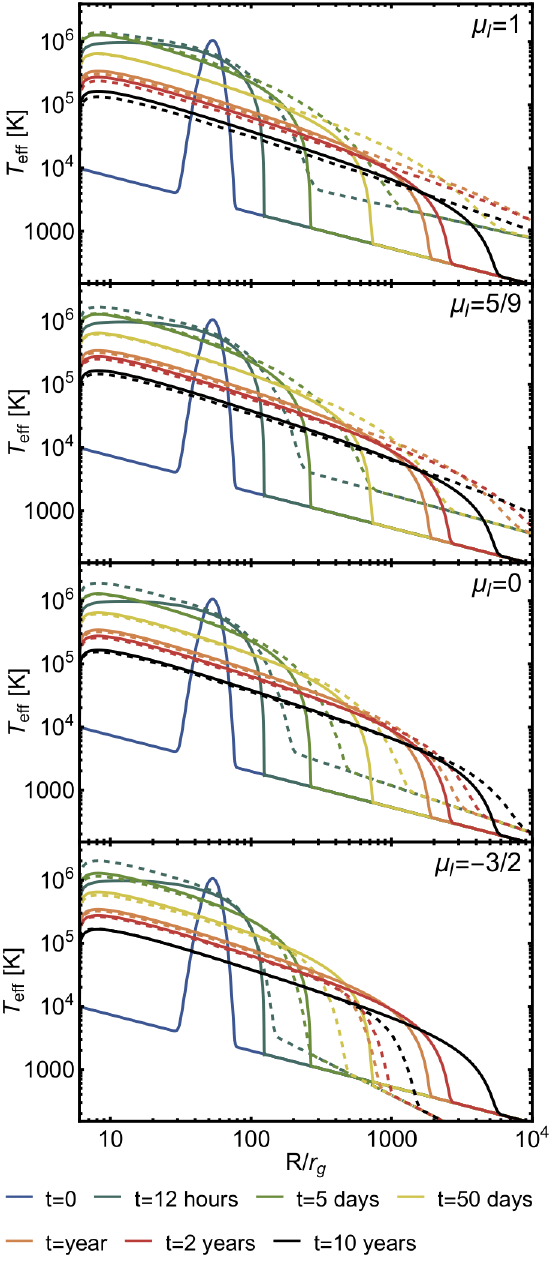}
\caption{The effective temperature $T_{\rm eff}$ vs radius at different times  for the magnetized viscosity (solid) and the linear viscosity
(dashed). The first case (\red{top}\blue{left}) assumes a Gaussian initial condition, where the normalization factor for the linear parametrization $\nu_{0,\rm l}$ is determined by matching the disk's outer radius after 1 year. The second case (\red{middle}\blue{right}) also assumes a Gaussian initial condition, but the normalization factor is chosen to match the magnetized viscosity at the initial condition. \red{The third case (bottom) includes a source term in the diffusion equation (Eq. \ref{eq:diffusion eq}) instead of using a Gaussian initial condition.} }
\label{fig: appendix TEff vs R}
\end{figure}

\begin{figure}
\includegraphics[width=85mm]{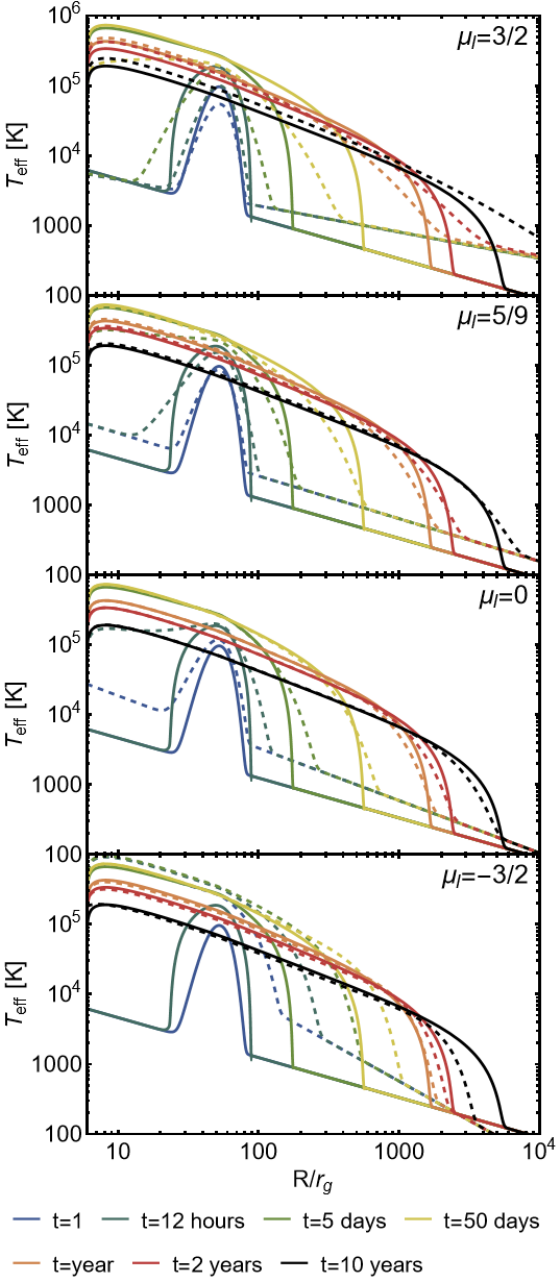}
\caption{\blue{Same as \cref{fig: appendix TEff vs R}, but includes a source term in the diffusion equation (Eq. \ref{eq:diffusion eq}) instead of using a Gaussian initial condition.} }
\label{fig: appendix TEff vs R source function}
\end{figure}

In \cref{fig: appendix Sigma vs R,fig: appendix light curves,fig: appendix TEff vs R,fig: appendix TEff vs R source function}, we explore these non-fiducial models by plotting the surface density and effective temperature as functions of disk radius, as well as the light curves. For all figures, we use same SMBH, star, and disk parameters as in \S\ref{sec: linear viscosity}.  The best match between the magnetized and linear viscosity models is achieved for $\mu_{\rm l}=0$ with the Gaussian initial condition normalized  at $1$ year, as discussed in the main text. 

\bibliography{main}{}
\bibliographystyle{aasjournal}

\end{document}